\documentclass[a4paper,12pt]{article}

\usepackage{amsmath,amssymb}
\usepackage{paralist}
\usepackage{slashed}
\usepackage{graphicx}
\usepackage{subfigure}
\usepackage{color}
\usepackage{fullpage}
\usepackage{multirow}
\usepackage{tikzfeynman}
\usepackage{hyperref}

\title{
\vspace{-3cm}
\begin{flushright}
\small{CERN-PH-TH/245}
\end{flushright}
\vspace{2.7cm}
\begin{center}
\bf \Huge
The 4D Composite Higgs
\end{center}
\vspace{.5cm}}
\date{}
\author{
{\large Stefania De Curtis$^{a,c}$\footnote{stefania.decurtis@fi.infn.it},~ Michele Redi$^{b,a}$\footnote{michele.redi@cern.ch},~
Andrea Tesi$^{c}$\footnote{andrea.tesi@yahoo.it}}\\
[10mm] \normalsize\itshape $^a$ INFN, Sezione di Firenze, Via G. Sansone, 1; I-50019 Sesto Fiorentino, Italy
\\
\normalsize\itshape
$^b$ CERN, Theory Division, CH-1211, Geneva 23, Switzerland\\
\normalsize\itshape
$^c$ Universit\`a degli Studi di Firenze, Dipartimento di Fisica e Astronomia,\\
\normalsize\itshape Via G. Sansone, 1; I-50019 Sesto Fiorentino,
Italy}

\begin{document}

\maketitle

\begin{abstract}
\noindent
We propose a four dimensional description of Composite Higgs Models which
represents a  complete framework for the physics of the Higgs as a pseudo-Nambu-Goldstone boson.
Our setup captures all the relevant features of 5D models and more in general of composite
Higgs models with partial compositeness. We focus on the minimal scenario where we include a single multiplet
of resonances of the composite sector, as these will be the only degrees of freedom which might be accessible at the LHC.
This turns out to be sufficient to compute the effective potential and derive phenomenological consequences of the theory.
Moreover our simplified approach is well adapted to simulate these models at the LHC.
We also consider the impact of non-minimal terms in the effective lagrangian which do not descend from a
5D theory and could be of phenomenological relevance, for example contributing to the $S-$parameter.
\end{abstract}

\newpage

\section{Introduction}\label{sec:1}

If the hierarchy problem is solved by strong dynamics we expect
resonances to appear around the TeV scale which will be hopefully within the reach of the LHC.
Producing these states would allow to determine precious informations about the
symmetries and dynamics of the strong sector. For example in technicolor theories the electro-weak
symmetry is spontaneously broken by strong dynamics according to the pattern
$SU(2)_L\times SU(2)_R/SU(2)_{L+R}$ and the dynamics of the techni-resonances
plays an important role.

Due to the notorious difficulties of theories without a Higgs, a logical possibility is that
the Higgs doublet itself is a composite state, as in this case the scale of strong dynamics could
be larger, alleviating phenomenological problems.  This can be most naturally realized
if the Higgs is a Goldstone Boson (GB) associated to the spontaneous breaking of a symmetry $G$ to $H$,
an idea which goes back to the '80s \cite{georgi}. One important ingredient of modern realizations is the mechanism of partial
compositeness where the SM fermions and gauge fields
mix to the composite states analogously to the photon-$\rho$ mixing in QCD.
The simplest example based on the symmetry pattern $SO(5)/SO(4)$ was considered
in \cite{Contino} in the context of Randall-Sundrum scenarios.
These models are also dual to strongly coupled 4D theories through the AdS/CFT correspondence \cite{ArkaniHamed:2000ds,Rattazzi:2000hs}.

In this paper we provide a general 4D effective description of Composite Higgs Models (CHM) with partial compositeness
(see \cite{continoreview} for a review and refs.).
Our starting point will be the $\sigma-$model $G/H$ describing the low energy interactions of GBs.
Spin-1 resonances non-linearly realizing the symmetry can be very simply included by introducing
an additional $\sigma-$model $G_{L}\times G_{R}/G_{L+R}$  and gauging the diagonal subgroup of $G_R$ and $G$.
Proceeding in the same  way a tower of resonances can be included at will. Fermionic resonances
can be treated in a similar way. Differently from other constructions we find it natural and economic
to include complete $G$ multiplets. Couplings to the elementary fields are simply accounted by
adding kinetic terms for the sources associated to the operators of the composite sector.

The motivation of our work is two-fold. Given that the 5D theories under consideration are effective theories
with a cut-off, it is natural to write an effective 4D lagrangian containing only the degrees of
freedom below the cut-off: up to cut-off dependent effects the two theories will coincide. This is particularly useful
because in phenomenologically relevant models only very few resonances lie below the cut-off.
The same idea was considered in \cite{adsqcd} in the context of 5D models for QCD.
Moreover only the lowest resonances have a chance to be produced
at the LHC, so for collider applications it is useful to write a truncation of the theory to the first multiplet of resonances.
One practical advantage is that in this limit very simple, explicit formulas can be derived which are
well suited for simulating the model at the LHC. Our philosophy is here similar to the one of \cite{contino-sundrum}.
In this ref. an effective lagrangian capturing the essential features of partial compositeness was considered but
the full GB structure of the theory was not incorporated. This was also done recently in \cite{panico-wulzer}
with a different construction from ours.

Since our framework is based on general symmetry principles it is also interesting to consider
a fully 4D point of view  and treat our effective theory with one set of $G$ multiplets on its own.
This can be interesting because terms not obtained from the 5D theory can be added to the action compatibly
with the symmetries which could be relevant phenomenologically. We will show in particular that they contribute
to the $S-$parameter allowing to reduce it for appropriate choice of the parameters.

The paper is organized as follows. In section \ref{sec:2} we develop the general formalism to include
an arbitrary tower of spin-1 and spin-1/2 resonances in a theory with global symmetry $G$ spontaneously broken to $H$.
Various limits are considered which reproduce in particular higher dimensional theories and the CCWZ construction \cite{CCWZ}.
In section \ref{sec:3} we apply our tools to the Minimal Composite Higgs $SO(5)/SO(4)$ focusing on the case where
a single $SO(5)$ multiplet of resonances is included. All the results can be conveniently expressed in terms of 2-point functions
of the composite sector which are collected in appendix \ref{sec:A}. The Higgs potential is discussed in section \ref{sec:4} and appendix \ref{sec:B}. In section \ref{sec:5} we study the effect of non-minimal terms in the effective action.
Conclusions are in section \ref{sec:6}. Couplings and decay widths of spin-1 resonances can be found in appendix \ref{appwidth}.

\section{Resonances in $G/H$}\label{sec:2}

We wish to describe scenarios where Higgs is GB of some strongly coupled sector with global symmetry $G$
spontaneously broken to a subgroup $H$. In the simplest realization \cite{Contino}, on which we will focus later on, the coset structure
is $SO(5)/SO(4)$ which delivers a single Higgs doublet. Our discussion however is general, relying solely on the symmetries
of the theory, and could be applied for example to extended Higgs sectors studied in \cite{singlet,cthdm}.

To lowest order in derivatives the dynamics of the GBs is determined by the symmetries.
In general, starting from the GB matrix\footnote{To fix our normalizations the generators are defined
with ${\rm Tr}[T^A T^B]=2 \delta^{AB}$.},
\begin{equation}
\label{gbmatrix}
U(\Pi)= e^{\frac {i \Pi^{\widehat{a}} T^{\widehat{a}}}f},
\end{equation}
which transforms under the action of the group $G$ as
\begin{equation}
\label{gbtransfor}
U(\Pi')= g U(\Pi) h^\dagger(\Pi,g) \ \ \ g\in G, \ \ \ h(\Pi,g) \in H,
\end{equation}
one constructs the Maurer-Cartan form $U^\dagger \partial_\mu U= i E^a_\mu T^a +i D^{\widehat{a}}_\mu T^{\widehat{a}}$,
where hatted and unhatted indexes correspond to broken and unbroken generators respectively. To leading order the
low energy GB lagrangian reads,
\begin{equation}
\frac {f^2} 2 D^{\widehat{a}}_\mu D^{\mu \widehat{a}},
\label{GBlagrangian}
\end{equation}
while the connection $E_\mu^a$ allows to write general $G$ invariant couplings to matter fields.
In the scenarios of interest, upon gauging the SM gauge interaction and introducing couplings
to the elementary SM fermions, eq.~(\ref{GBlagrangian}) describes the low energy dynamics of CHM with partial compositeness.
The GB structure has important consequences on the couplings of the Higgs which was discussed
in a model independent fashion in \cite{SILH}. In particular this encodes the relevant low energy consequences
of 5D models such as \cite{Contino}.

To add resonances to this picture we extend the construction of \cite{sonstephanov}, which was formulated in the context of QCD. See also \cite{thaler} for
a related construction.
\begin{figure}[t]
\begin{center}
\includegraphics[width=\textwidth]{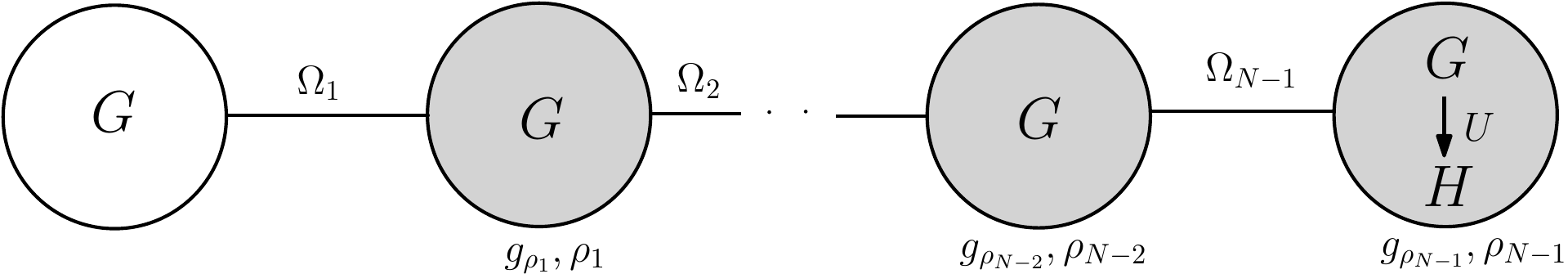}
\caption{\label{fig:1}\small $Moose$ diagram of the theory. The white site on the left corresponds to the global $G$ symmetry, while the grey ones are gauged.}
\end{center}
\end{figure}

\subsection{Spin 1}\label{sec:2.1}

Starting from the GB lagrangian (\ref{GBlagrangian}), a simple way to add vector resonances compatibly with the symmetries
is to introduce a second $\sigma-$model field $\Omega$ of $G_L \times G_R/ G_{L+R}$. This can be parametrized
by $G$ valued matrices transforming as
\begin{equation}
\label{transform}
\Omega \rightarrow  g_L \Omega g_{R}^\dagger.
\end{equation}
A lagrangian manifestly invariant under a global $G'$ symmetry spontaneously broken to $H$ containing spin 1 resonances
($\rho_\mu$) is obtained by gauging the diagonal subgroup of $G_R$ and $G$ in eq.~(\ref{GBlagrangian}),
\begin{equation}
\label{gauge2site}
{\cal L}_{2-site}=\frac{f_1^2}{4}{\rm Tr}\left|D_{\mu}\Omega \right|^2 +  \frac {f_2^2} 2 {\cal D}^{\widehat{a}}_\mu {\cal D}^{\mu \widehat{a}}-\frac 1 {4 g_\rho^2}
\rho_{\mu\nu}^A\rho^{A\mu\nu},
\end{equation}
where
\begin{equation}
\label{der_cov}
D_{\mu}\Omega=\partial_{\mu}\Omega-i A_{\mu} \Omega+i \Omega \rho_{\mu},
\end{equation}
${\cal D}^{\widehat{a}}$ is obtained by gauging the global symmetry of the $\sigma-$model (\ref{GBlagrangian}),
and $\rho_{\mu\nu}^A$ is the field strength of $\rho_\mu^A$.
We have also included in the covariant derivative (\ref{der_cov}) the gauging of $G_L$. From the point of view of the
composite sector this is a global symmetry so that $A_{\mu}$ is not a dynamical field and should
be treated as an external classical source which is zero in the vacuum.
These sources are useful to compute correlation functions of the global currents the theory: the path integral as a function of $A$ is the
generating functional of the conserved currents.  Moreover, as we will see in the next section,
the coupling  to external fields (corresponding to SM fermions and gauge fields) is simply accounted by adding kinetic
terms for the sources of the various fields. One can then determine the effects of the composite sector on
the SM fields in terms of the correlation functions of the composite sector.

This procedure can be generalized in an obvious way to add a tower of resonances by
adding $\sigma-$model fields $\Omega_n$ and gauging nearest neighbor diagonal groups.
One finds the moose lagrangian  depicted in figure \ref{fig:1},
\begin{eqnarray}
\label{gauge}
&&{\cal L}_{N-sites}=\sum_{n=1}^{N-1} \frac{f_n^2}{4}{\rm
Tr}\left|D_{\mu}\Omega_n \right|^2   +
\frac{f_{N}^2}{2}{\cal D}^{\widehat{a}}_\mu {\cal D}^{\mu \widehat{a}}
-\sum_{n=1}^{N-1}\ \frac{1}{4\,g_{\rho_n}^2} \rho_{n,\mu\nu}^A\rho_n^{A\mu\nu}\nonumber \\
&&D^{\mu}\Omega_n=\partial^{\mu}\Omega_n-i \rho^{\mu}_{n-1}\Omega_n+i \Omega_n \rho^{\mu}_n,
\ \ \ \ n=1,..,N-1
\end{eqnarray}
where $\rho_n^{\mu}\in Adj[G_n]$ and $\rho_0^\mu= A^\mu$. The spectrum now contains $N-1$ $G$ multiplets of
massive spin-1 resonances. The breaking $G/H$ is triggered by the last $\sigma-$model.
The GBs associated to the breaking $G/H$ can be identified with the combination,
\begin{equation}
\label{GBfield}
\boldsymbol{U'}\equiv \left(\Pi_{n=1}^{N-1} \Omega_n \right) U.
\end{equation}
To make manifest the particle content of the theory it is useful to choose a gauge where
the GBs do not mix with the heavy gauge bosons. This is given by
\begin{equation}
\label{unitarygauge}
\Omega_n=\exp i\frac{f}{f_n^2}\Pi, \ \ \ n=1,...,N
\end{equation}
with $U=\Omega_N$. The scale $f$ can be conveniently chosen to be the decay constant of the GBs in the theory.
From the nomalization of the kinetic terms one quickly derives,
\begin{equation}
\sum_{n=1}^N\frac{1}{f_n^2}= \frac{1}{f^2}.
\label{fpi}
\end{equation}

\subsection{Spin $\frac 1 2$}
\label{sec:2.2}

Fermions can be treated in a similar way with one subtlety. The simplest option is to assume that the
composite sector contains Dirac fermions arising from the strong dynamics. Each SM chiral fermion
will be associated to one (or more) $G$ multiplets of the strong sector.
We will then add Dirac fermions $\Psi_n$ in a representation $G_n$ (same at each site), in general reducible.

\begin{figure}[t]
\begin{center}
\includegraphics[width=0.7\textwidth]{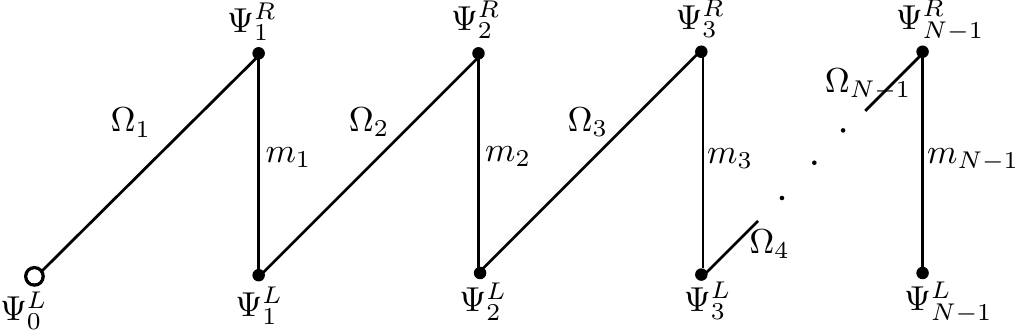}
\caption{\label{fig:2}\small Nearest neighbor interactions in the fermion sector. For fields associated to left-handed sources, the left chirality at site $n$ couples to the right chirality at site $n+1$.}
\end{center}
\end{figure}

As for the gauge field we can write a lagrangian with nearest neighbor interactions.
For the states associated to the left-handed SM chiralities we will consider,
\begin{eqnarray}
\label{fermions}
{\cal L}_{fermions} &=& \sum_{n=1}^{N-1}
\bar{\Psi}^{(r)}_n \left[i \slashed{D}^{\rho_n} -m_n^{(r)}\right]\Psi^{(r)}_n+\sum_{n=1}^{N-1}{\Delta^{(r)}_n}
\left(\ \bar{\Psi}_{r,L}^{n-1}\Omega_n
\Psi_{r,R}^{n} + h.c.\right),\nonumber\\
D^{\mu}\Psi^{(r)}_n &=& \partial^\mu\Psi^{(r)}_n-i\rho^\mu_n\Psi^{(r)}_n,
\end{eqnarray}
while for right-handed fermions $L\leftrightarrow R$. The index $r$ refers to the irreducible representations of $G$ and summation is understood.
The hopping terms written above are not the most general ones, connecting only the left chirality of the massive fermion at site $n$ to the right
chirality of the fermion at site $n+1$, as shown in figure \ref{fig:2}. This choice can be motivated as follows. As in the gauge sector the field
on the first site $\Psi^{(r)}_0$ is a classical source. Differently however, we would like the associated operator to be chiral so
that only one chirality of $\Psi^{(r)}_0$ is identified as the source. This suggests to interpret at each site the left (right) fields as
sources for the fields on the next site, while the right-handed (left-handed) fermions as independent dynamical fields.
This leads to the lagrangian above.

Following the same logic, on the last site we write the most general zero  derivatives lagrangian compatible
with the spontaneously broken symmetry and with the $LR$ structure.
As explained in \cite{cthdm} this is achieved by writing all the possible $H$ invariant terms constructed with GBs,
\begin{equation}
\label{yuk}
{\cal L}_{\frac G H} = m_\Psi \sum_{r,s} \bar{\Psi}^{(r),N-1}_{L}U(\Pi) P^{rs}_A U(\Pi)^\dagger {\Psi}^{(s),N-1}_{R} + h.c.
\end{equation}
where $P_A^{rs}$ is a projector over the irreducible representations, $r$ refers to the fields associated to the left sources
and $s$ to the ones associated to the right-sources. This formalism also allows to introduce composite massless fermions
in a representation $H$ of the symmetry  which will interact with the fields on the last site by means of the GBs.

\subsection{Other Terms}
\label{sec:2.3}

The effective lagrangians (\ref{gauge}), (\ref{fermions}) and (\ref{yuk}) are not the most general compatible with the symmetries
of the theory and this provides powerful restrictions on the spectrum and the dynamics of the theory. Consider the
case with only one multiplet of resonances. To the gauge lagrangian in eq.~(\ref{gauge2site}) we can add the term,
\begin{equation}
\frac {f_0^2} 2 \tilde{D}^{\widehat{a}}_\mu \tilde{D}^{\mu \widehat{a}},
\label{f0term}
\end{equation}
where $\tilde{D}^{\mu \widehat{a}}$ is now obtained from the Maurer-Cartan form for $\Omega U$.
The addition of this term changes the GBs decay constant to
\begin{equation}
f^2= f_0^2 + \frac {f_1^2 f_2^2}{f_1^2+f_2^2},
\label{fhls}
\end{equation}
while leaving the masses of composite gauge bosons unchanged.
One important consequence is that the coupling of spin-1 resonances to GBs is also modified.

While this term violates the logic of nearest neighbor interactions, there are several reasons to include it.
In a phenomenological approach to QCD it is needed to better reproduce the data \cite{bando}. In that case one finds that the
best agreement is for $f_0^2\approx-f_\pi^2$. Moreover if we imagine starting
from a theory with many sites, we could integrate out the heavier states and write an effective
lagrangian for the lighter ones. This term would then be induced at tree level.
Finally we expect these terms, appropriately suppressed, to be generated by quantum loops.

From a 5D point of view (see below) the term above corresponds to a non-local interaction.
This should be suppressed if the 5D theory is weakly coupled. Indeed one can see that
such term, if large, tends to ruin the partial unitarization of the scattering of GBs that one achieves in
a theory with nearest neighbor interactions (see below). However phenomenologically interesting
theories require strong coupling,  so that just few resonances have a weakly coupled description.
In this situation it is interesting to consider phenomenological consequences of (\ref{f0term}).
As we will see in section \ref{sec:5} this term could be relevant for the decay of the resonances and contributes
to the $S$ parameter. Analogous terms could also be introduced for the fermions but we will not consider their effects in this paper.

\subsection{Relation to Other Works}\label{sec:2.4}

\subsubsection{5D Models}

The  reader will likely not have missed that the lagrangians (\ref{gauge}) and (\ref{fermions}) resemble the discretization
of a 5D theory. Our approach indeed is reminiscent of dimensional deconstruction \cite{deconstruction}.
We wish to make the connection manifest. The discussion will follow the one in
\cite{adsqcd} in the context of extra-dimensional models for QCD.
Let us first consider the gauge sector. The continuum limit is obtained by taking the number of sites $N\to \infty$ with
\begin{equation}
\begin{split}
& f_n^2 = \frac{a[(z_{n-1}+z_n)/2]}{g_5^2}\frac{2\,N}{L}, \\
& g_n^2 = \frac{g_5^2}{a(z_n)}\frac{N}{L},
\end{split}
\end{equation}
where $g_5$ and $L$ are identified respectively the extra-dimensional gauge coupling and interval length. $a(z)$ is a smooth function and $z_n=z_0+n L/N$. One can check that the lagrangian (\ref{gauge}) reproduces a gauge theory in a space with metric $g_{MN}=a(z)^2 \eta_{MN}$.
Choosing couplings and decay constants an arbitrary metric can be obtained.

This picture however is only true at the classical level. Quantum mechanically the 5D gauge theory
is a non-renormalizable effective theory meaningful only at energies below the 5D cut-off,
\begin{equation}
\Lambda_5 \sim \frac {16\pi^2}{g_5^2}.
\end{equation}
As a consequence only the resonances below this energy scale are weakly coupled and can be trusted within
the effective description while the heavier states parametrize cut-off dependent effects.
The 4D theory is also an effective theory where the cut-off originates from the non-linearities of the $\sigma-$models.
In a theory with just GBs the maximum cut-off would be $4\pi f$ (corresponding to
the scale where the scattering of GB becomes strongly coupled) which is in general smaller than
$\Lambda_5$. In our lagrangian with nearest neighbor interactions the local cut-off is given by $4\pi f_n$.
From (\ref{fpi}) $f_n$ grows with the number of links. The cut-off is than larger that the naive cut-off and indeed one can show that
the 5D one can be reproduced. The reason for this is that the resonances partially unitarize the scattering of GBs
allowing the theory to be reliable up to energies parametrically higher than $4\pi f $. In practice phenomenologically
interesting theories demand the effective coupling of the higher dimensional theory to be rather strong
so that just a few resonances will be captured in an effective description. In this sense the notion of
an extra-dimension becomes rather ethereal as there is no regime where five-dimensional physics is recovered.
In this situation it is natural to think of this theory as a 4D theory with just a few states.

Let us turn to the spontaneous breaking of the symmetry.
In the language of the 5D theory the last site corresponds to the IR brane where the symmetry $G$ is spontaneously
broken  to $H$. This is often achieved imposing Neumann or Dirichlet boundary  conditions for the gauge fields
associated to the broken or unbroken generators respectively. In the language of the deconstructed theory this corresponds
to setting $f_N\to \infty$. Such an infinite VEV however is unlikely to be realized physically. A more
physical picture is that some dynamics is responsible for the breaking of the symmetry in the IR producing a finite $f_N$.
We will leave  $f_N$ as a free parameter in what follows.

In the fermion sector the discretization of the 5D theory is less trivial. Usually theories obtained by naive discretization
suffer from fermion doubling and need to be regulated by adding term of higher order in derivatives relevant for the
deconstructed theory and irrelevant for continuum theory (see for example \cite{Hill:2002me}). One can show
that our fermionic action (\ref{fermions}) has the correct continuum limit and this is in fact another motivation for our choice.

\subsubsection{SILH and Discrete Models}

In the ``Strongly Interacting Light Higgs'' (SILH) \cite{SILH}, the effective lagrangian for the Higgs as a GB was studied,
focusing on the low energy consequences of the symmetry. It can be seen that the leading order lagrangian in an expansion
$E/m_\rho$ corresponds to our setup with one site (no composite resonances) and kinetic term for the sources corresponding
to the SM fields. In the model with two or more sites integrating out the heavy resonances will generate higher order terms in $E/m_\rho$
of  phenomenological importance in the low energy action, reproducing sub-leading terms with the appropriate power counting described in \cite{SILH}.  Our setup can then be considered as the extension of SILH with resonances integrated in.
This is of course mandatory to study the production of these states.

A standard way to introduce spin 1 resonances into this picture is through the CCWZ formalism, see \cite{Contino:2011np}
for a recent study within $SO(5)/SO(4)$. For spin 1 resonances in the adjoint of $H$ this can be realized by considering
gauge fields of $H$ with a lagrangian,
\begin{equation}
{\cal L}=\frac {f^2}{2} D_\mu^{\widehat{a}} D^{\mu \widehat{a}}-\frac 1 {4 g_\rho^2} \rho_{\mu\nu}^a \rho^{\mu\nu a} +\frac {f'^2} {2} (\rho_\mu^a - E^a_\mu)^2.
\label{CCWZ}
\end{equation}
Note that spin 1 resonances in a different representation of $H$ cannot be coupled in this way, since gauge fields
are necessarily in the adjoint of the group. As we will see, the resonances in $G/H$ can play an important role.
For matter fields instead any $H$ invariant lagrangian can be lifted to a $G$ invariant one with the aid of
the GBs.

In our construction fields are complete $G$ multiplets so that we treat resonances in $H$ and $G/H$ on the same ground.
This is quite natural in QCD where mesons fill complete representations of the chiral symmetry.
We can however recover the CCWZ lagrangian with only $H$ resonances starting from the
model with one $G$ multiplet and taking $f_2\to \infty$ in eq.~(\ref{gauge2site}). In this limit the spin-1 resonances in $G/H$
become infinitely heavy and one can easily check that one obtains the lagrangian (\ref{CCWZ}).
The non-nearest neighbor  interaction, introduced in section \ref{sec:2.3}, is necessary to reproduce the most general
lagrangian (\ref{CCWZ}) where $f$ and $f'$ are different.

For the case of two sites with symmetry structure $SO(5)/SO(4)$, on which we will focus in the rest of the paper,
our construction is philosophically similar to ref.~\cite{contino-sundrum}, where however
the non-linear GB structure was not included. Roughly the lagrangian in \cite{contino-sundrum} can
be understood as the truncation of the GB lagrangian to the lowest dimension operators.
One difference is however that the Higgs necessarily couples to the elementary fields in our case.
The extension of ref.~\cite{contino-sundrum} to the GB Higgs was also considered in ref.~\cite{panico-wulzer}
where a discrete model to describe composite resonances was proposed based on
$G_L\times G_R/G_{L+R}$ $\sigma-$models\footnote{A two site description can be also found in \cite{Barb-Bell}. Our gauge sector is similar to the one described in section 6.2 of that work with the scale $F$ kept finite. A ``Little-Higgs'' inspired 2-site model is described also in \cite{Foadi}.}.
In that work the spontaneous breaking  $G/H$ was realized by gauging a subgroup $H$ of the $G_R$ symmetry of
the last $\sigma-$model $G_L\times G_R/G_{L+R}$. While this breaks explicitly the symmetry $G_R$
on the last site, it can be seen that the action describes the spontaneous breaking $G/H$ and contains incomplete $G$ multiplets.
A different choice was also considered for the fermionic fields. In our construction we find it natural to start
from the $\sigma-$model $G/H$ and add complete $G$ multiplets of resonances to this picture as
explained above, effectively extending \cite{SILH}. For the gauge sector the setup of \cite{panico-wulzer} with $N$ sites
can be recovered starting from our $N-$sites lagrangian in the limit $f_N\to \infty$.

\section{Minimal 4D Composite Higgs}
\label{sec:3}

\begin{figure}[t]
\begin{center}
\includegraphics[width=\textwidth]{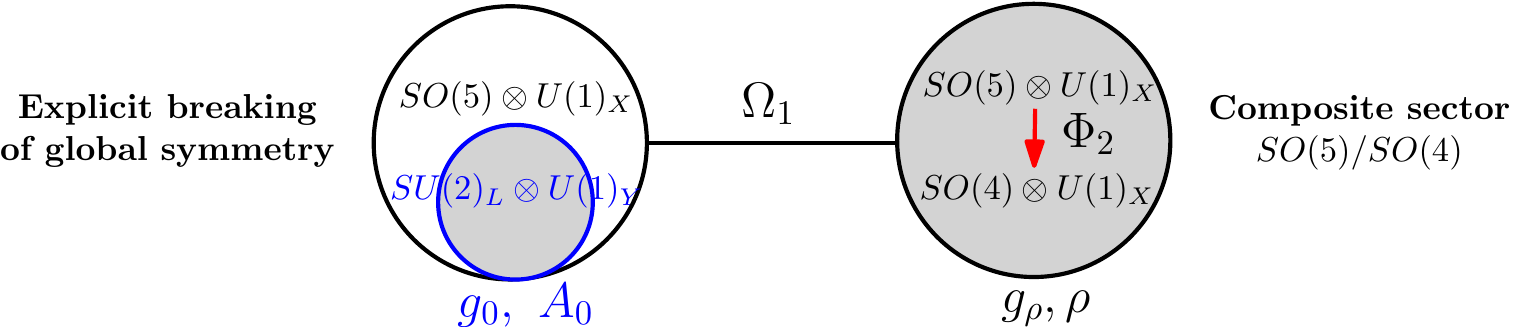}
\caption{\label{fig:2-site}\small 2-site model $SO(5)/SO(4)$: gauge sector.
The first site  is the elementary sector, the second the composite sector with $SO(5)\times U(1)_X$ heavy
multiplets.}
\end{center}
\end{figure}

In the previous section we discussed the features of the composite sector relevant to build CHM where the Higgs is a GB.
We will now apply this framework to the Minimal Composite Higgs of ref.~\cite{Contino}.
As in that work, we find it very useful to express our results in terms of the correlation functions of the composite sector
which we collect in appendix \ref{sec:A}.  We will focus on the case with two sites depicted
figure \ref{fig:2-site},  even though the formulas in terms of the correlation functions are general.
There are two main reasons for this choice. Most importantly the 2-site lagrangian contains only the states which might
be accessible at the LHC capturing the relevant features of all CHM with partial compositeness.
Secondly this truncation allows to obtain extremely simple formulas depending on few parameters
which are ready made to simulate the scenario at the LHC.

The symmetry structure of the theory is $SO(5)/SO(4)$ which delivers 4 GBs
in the vector representation of $SO(4)$, appropriate for the Higgs boson.
The breaking can be parametrized by the VEV of a vector of $SO(5)$
so that the gauge lagrangian (\ref{gauge2site}) can be written explicitly as\footnote{
We follow the conventions of \cite{Contino}. In particular we have redefined the generators so that in the vector representation
${\rm Tr}[T^A T^B]=\delta^{AB}$. This is convenient as the $SU(2)_L$ subgroup has then standard structure constants $\epsilon_{ijk}$.},
\begin{equation}
{\cal L}_{gauge}=\frac{f_1^2}{4}{\rm Tr}\left|D_{\mu}\Omega_1 \right|^2 +  \frac {f_2^2} 2 \left(D_{\mu}\Phi_2\right)\left(D^{\mu}\Phi_2\right)^T -\frac 1 {4 g_\rho^2}
\rho_{\mu\nu}^A\rho^{A\mu\nu},
\label{gaugeminimal}
\end{equation}
where $|\Phi_2|^2=1$.

The GBs can be identified as explained in section \ref{sec:2.1}. Introducing the GB matrix,
\begin{equation}
\label{Definizione_PAI}
\Pi=\sqrt{2}h^{\widehat{a}}T^{\widehat{a}}=-i\left(\begin{array}{cc} 0_4 & {\bf h} \\ -{\bf h}^T & 0\end{array}\right)
\end{equation}
and with the parametrization (\ref{unitarygauge}) we find
\begin{equation}
\label{omega_n}
\Omega_n = {\bf 1} +i\frac{s_n}{h}\Pi + \frac{c_n-1}{h^2}\Pi^2,\ \ \ \ s_n\equiv \sin \left(f h/ f^2_n \right), \ \ c_n\equiv \cos \left(f h/ f^2_n \right), \ \ h\equiv\sqrt{h^{\widehat{a}}h^{\widehat{a}}}
\end{equation}
with $n=1,2$. We define $\Phi_2=\phi_0\Omega_2^T$, where $\phi_0^i=\delta^{i5}$.
Following eq.~(\ref{GBfield}), the GBs are parametrized by,
\begin{equation}
\label{GBSO5}
\Phi=\Phi_2\Omega_1^T= \phi_0 e^{ -i \frac{\Pi}{f}}=\frac {1} {h}\sin \frac{h}{f}\left( h_1,\,h_2,\,h_3,\,h_4,\,h\,\cot \frac{h}{f}\right).
\end{equation}

To reproduce hyper-charge assignments of the SM fermions we need to extend the symmetry
of the composite sector to $SO(5)\times U(1)_X$  where the latter symmetry is unbroken in the
vacuum. The hyper-charge is identified with the combination
\begin{equation}
Y= T_{3R}+X.
\end{equation}
In the framework of partial compositeness the composite sector should also possess
an $SU(3)$ global symmetry, corresponding to color. However this does not play a role
in what follows so we will omit it (except for taking into account the color factors).

Having discussed the composite sector, the next step is to introduce SM fermions and gauge fields.
Practically this amounts to adding kinetic terms for the sources of the operators of the composite sector
to which the SM field couple and setting to zero non-dynamical fields.
To gauge the SM gauge symmetry we then introduce the kinetic terms,
\begin{equation}
\label{ele-gauge}
{\cal L}_{gauge}^{el}=-\frac 1 {4 g^2_{0}} F^a_{\mu\nu} F^a_{\mu\nu}-\frac 1 {4 g^2_{0Y}} Y_{\mu\nu} Y^{\mu\nu}
\end{equation}
for the sources associated to the $SU(2)_L$ and $U(1)_Y$ symmetries. Formally non-dynamical fields correspond
to infinite elementary kinetic terms.

For fermions it works in a similar way. We add kinetic terms for the elementary SM fermions (we will focus here only on the third gerneration quarks),
\begin{equation}
\label{ele-ferm}
{\cal L}_{fermions}^{el}=\frac{1}{y_{q_L}^2} \bar{q}^{el}_L i \slashed{D}^{el} q^{el}_L + \frac{1}{y_{t_R}^2} \bar{t}^{el}_R i \slashed{D}^{el} t^{el}_R +\frac{1}{y_{b_R}^2} \bar{b}^{el}_R i \slashed{D}^{el} b^{el}_R,
\end{equation}
where $\slashed{D}^{el}$ corresponds to the covariant derivative with respect to the elementary fields.
There exists also the possibility that one chirality, typically the right-handed top, is part of the strong sector in which case
no elementary component exists but we will not consider it here.
To determine the model we just need to specify to which sources the elementary fields correspond.
This in general depends on the fermionic representation of the composite sector.

The elementary lagrangian explicitly breaks the global symmetry of the theory so that the GBs become approximate.
In particular the VEV becomes physical and the vacuum alignment determines the electro-weak VEV.
The main consequences are the generation of mass terms for $W^\pm$ and $Z$ and Yukawa couplings for SM fermion.
The explicit breaking also generates a potential for Higgs at 1-loop which we investigate in section \ref{sec:4}.

As an example we consider in detail the model where SM fermions couple to {\bf 5} reps of $SO(5)$ as in \cite{custodian}.
This is a realistic model which allows to satisfy precision electro-weak measurements as well as generating successful electro-weak symmetry breaking (EWSB).
A similar analysis can be done for other models and it is essentially
determined once the representation of the composite sector are fixed.

We consider first the gauge sector which is model independent.

\subsection{Gauge Sector}\label{sec:3.1}

Composite spin 1 resonances fill an adjoint representation of $SO(5)\times U(1)_X$.
Before coupling to the elementary gauge fields the spectrum has an unbroken $SO(4)\times U(1)_X$ symmetry,
\begin{equation}
\begin{aligned}
 m^2_{\rho}&= \frac{g_\rho^2 f^2_1 }{2}, \\
 m^2_{a_1}&= \frac{g_\rho^2(f_1^2+f_2^2)}{2},\\
 m_{\rho_X}^2&=\frac {g_{\rho_X}^2 f_X^2}{2},
\end{aligned}
\label{masses1}
\end{equation}
where in analogy with QCD we denoted the masses of the resonances in $H$ and $G/H$ as $m_\rho$ and $m_{a_1}$ respectively
(see \cite{Agashe:2009bb} for a related discussion).
The mass and coupling of the resonance associated to $U(1)_X$ can be different from $m_\rho$ so $g_{\rho_X}$ and $f_X$
are in general independent parameters. For simplicity we will take $f_X=f_1$ in what follows.
These formulas continue to hold also when $f_0\ne 0$. While from eq.~(\ref{fhls})
for $f_0=0$ the sign of $f_2^2$ is strictly positive, it could in principle be negative for $f_0\ne 0$ leading
to coset resonances lighter than $H$ resonances.

The kinetic terms for the SM gauge fields breaks the $SO(4)$ degeneracy. The fields decompose into $SU(2)_L\times U(1)_Y$ multiplets
with masses given by the zeros of eqs.~(\ref{Pi_0}). The mass of composite states coupled to the elementary fields are in
general heavier than their $SO(4)$ partners.
For example for $SU(2)_L$ we have one massless triplet of gauge fields ($W^{aL}$) and one heavy
triplet of fields with mass:
\begin{equation}
\label{massL}
m_{\rho_{aL}}= \frac{m_\rho}{\cos\theta_L},\qquad \tan\theta_L = \frac {g_0}{g_\rho}.
\end{equation}

From the correlation functions in appendix \ref{sec:A} we can identify the low energy parameters of the theory.
The electro-weak gauge couplings are given by,
\begin{equation}
\label{couplingL}
\begin{aligned}
\frac{1}{g^2}&= -\Pi_0'(0) = \frac{1}{g_0^2}+\frac{1}{g_\rho^2},\\
\frac{1}{g'^2}&= -\Pi_Y '(0) = \frac{1}{g_{0Y}^2}+\frac{1}{g_\rho^2}+\frac{1}{g_{\rho_X}^2},
\end{aligned}
\end{equation}
while the electro-weak VEV is
\begin{equation}
\label{VEV}
v= f \sin \frac{\langle h \rangle}{f}.
\end{equation}

The interaction vertices and decay of the resonances are studied in appendix \ref{appwidth}.

\subsection{Fermions}\label{sec:3.2}

As an example we consider the first model of ref.~\cite{custodian} where SM fermions couple to fermionic operators in the {\bf 5} of $SO(5)$.
This model, known as CHM$_5$, is a realistic scenario compatible with precision electro-weak measurements.
Analogous results can be derived for any other fermionic choice, see appendix \ref{sec:A} for the model with {\bf 10} of $SO(5)$ (CHM$_{10}$).
For simplicity we will consider only the third generation quarks, which will be relevant for the computation of the potential in the next section.

\begin{figure}[t]
\label{fig:CHM5}
\begin{center}
\includegraphics{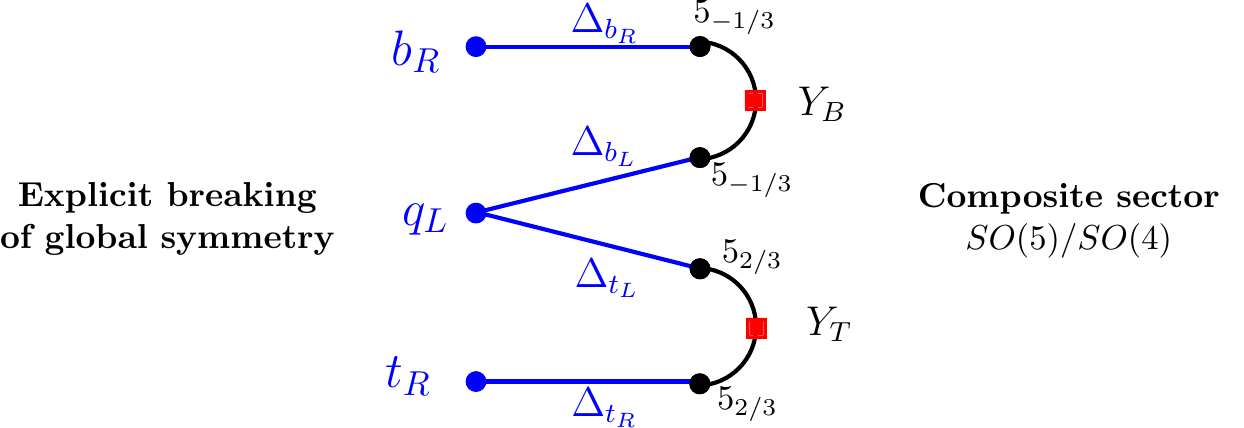}
\caption{\small 2-site model $SO(5)/SO(4)$: fermionic sector. The spontaneous breaking of $SO(5)$
is achieved through Yukawa couplings (\ref{yuk}) in the composite sector, here drawn as red squares.}
\end{center}
\end{figure}

Under $SU(2)_L\times SU(2)_R$ the vector representation of $SO(5)$ decomposes as a $({\bf 2},{\bf 2})$ and a $({\bf 1},{\bf 1})$.
The SM left doublet can be embedded in a
$({\bf 2},{\bf 2})_{2/3}\in \Psi_T$ as,
\begin{equation}
{\bf 5}_{2/3} = ({\bf 2},{\bf 2})_{2/3} \oplus ({\bf 1},{\bf 1})_{2/3},\qquad ({\bf 2},{\bf 2})_{2/3}=\left(\begin{array}{cc}
T & T_{\frac 5 3} \\
B & T_{\frac 2 3}
\end{array}\right)
\label{rep1}
\end{equation}
while $t_R$ can be coupled to a singlet in a different ${\bf 5}$ rep, $\Psi_{\widetilde T}$.
For the bottom sector $b_R$ is coupled to the singlet in a ${\bf 5}_{-1/3}$ ($\Psi_{\widetilde{B}}$)
so to generate the bottom Yukawa it is also necessary (by $U(1)_X$ symmetry) to couple the SM doublet to a second
doublet in ${\bf 5}_{-1/3}$ ($\Psi_{B}$) which contains
\begin{equation}
{\bf 5}_{-1/3} = ({\bf 2},{\bf 2})_{-1/3} \oplus ({\bf 1},{\bf 1})_{-1/3},\qquad ({\bf 2},{\bf 2})_{-1/3}=\left(\begin{array}{cc}
B_{-\frac 1 3 } & T'  \\
B_{-\frac 4 3 } & B'
\end{array}\right).
\end{equation}
To summarize, the spectrum contains ${\bf 5}_{2/3}$ ($\Psi_T,\Psi_{\widetilde{T}}$) and
${\bf 5}_{-1/3}$ ($\Psi_B,\Psi_{\widetilde{B}}$) reps, associated respectively to the top and bottom sectors.
This choice of representations, while minimal, has special phenomenological virtues suppressing
dangerous corrections to the couplings of the $Z$ which otherwise would be  problematic phenomenologically \cite{zbb}.

Following section \ref{sec:2}, the effective lagrangian for the composite states and elementary fields reads\footnote{To introduce the Yukawas of the down sector we define the source of the quark
doublet as $\Psi_L=\cos\varphi \Psi^0_T + \sin\varphi \Psi^0_B$ and the orthogonal combination
$\Psi^\perp_L=-\sin\varphi \Psi^0_T + \cos\varphi \Psi^0_B$. We then identify
$y_{t_L}=y_{q_L}\cos\varphi$ and $y_{b_L}=y_{q_L}\sin\varphi$. In order to reproduce $m_b\ll m_t$ it is natural to take
$\sin\varphi \ll 1$  so that the mixing in the down sector is small and can be neglected for some purposes such as computation of
the potential. Moreover this choice  guarantees that the shift of the coupling of $b_L$ to the $Z$ is small.}
\begin{equation}
\label{CHM5}
\begin{aligned}
{\cal L}^{\rm CHM_5} &= {\cal L}_{fermions}^{el} \\
&+ \Delta\,  \bar{q}^{el}_L \Omega_1 \Psi_{T}  +\Delta\, \bar{t}^{el}_R \Omega_1 \Psi_{\widetilde{T}} +  h.c. \\
 & + \bar{\Psi}_T (i \slashed{D}^{\rho} -m_{T}) \Psi_T +  \bar{\Psi}_{\widetilde{T}} (i \slashed{D}^{\rho} -m_{\widetilde{T}}) \Psi_{\widetilde{T}}     \\
 & -  Y_T \bar{\Psi}_{T,L} \Phi_2^T \Phi_2 \Psi_{{\widetilde{T}},R}- m_{Y_T}\bar{\Psi}_{T,L} \Psi_{{\widetilde{T}},R}+h.c.\,  \\
 & + (T\rightarrow B),
\end{aligned}
\end{equation}
where $\Delta$ is introduced for dimensional reasons.
As explained in section \ref{sec:2.2} we do not include all the possible terms in the composite sector allowed by the symmetries
but only the ones with the required $LR$ structure. This is actually the minimal choice to generate the SM Yukawas.
The $SO(5)$ invariant proportional to $m_{Y_T}$ is needed to describe the most general embedding of the elementary fields.

To leading order the mass of the top is given by,
\begin{equation}
\label{top-approx}
m_t \sim \frac v {\sqrt{2}}
\frac {\Delta_{t_L}} {m_T}
\frac {\Delta_{t_R}} {m_{\widetilde T}}
 \frac {Y_T} f
\end{equation}
where we defined,
\begin{equation}
\label{mixingD}
\Delta_{t_L} = y_{t_L} \Delta,\qquad \Delta_{t_R}=y_{t_R} \Delta.
\end{equation}
It is natural to interpret $Y_T/f$ as the coupling of the composite sector and $\Delta_{t_L}/m_T$, $\Delta_{t_R}/m_{\widetilde T}$
as the elementary-composite mixings. We should mention that the approximation above, excellent for the light quarks,
is not always sufficient for the top, the exact expression used in our numerical computations is reported in eq.~(\ref{massatop}).

The mass spectrum is in general most easily extracted from the fermionic correlators in appendix \ref{sec:A}.
In the top sector breaking of $SO(4)$ symmetry due to the mixings is sizable
so the spectrum is only (approximately) $SU(2)_L\times U(1)_Y$ symmetric.
Before EWSB the spectrum of heavy fermions is given by,
\begin{itemize}
\item  two ${\bf 2}_{1/6}$, zeros of $\Pi^{q}_0$,
\item  two ${\bf 2}_{7/6}$, poles of $\widehat{\Pi}^{q_L}_0$,
\item  two ${\bf 1}_{2/3}$, zeros of $\Pi^{u}_0$,
\end{itemize}
where form factors are defined in (\ref{SE_5y}, \ref{CHM5-Piq}). Rather than writing the explicit formulas, computable as explained,
in the next section we will present the fermionic spectrum as a function of the Higgs mass.
Here we only anticipate that top partners inside ${\bf 2}_{7/6}$ can be very light, in the region of large elementary-composite mixings (\ref{mixingD}),
\begin{equation}
m_{{\bf 2}_{7/6}}^2 = \frac{1}{2}\left[ m_T^2 + m_{\widetilde T}^2 + m_{Y_T}^2 \pm \sqrt{(m_T^2-m_{\widetilde T}^2)^2+m_{Y_T}^2(m_{Y_T}^2+2(m_T^2+m_{\widetilde T}^2))}~\right].
\end{equation}

\section{Higgs Potential}
\label{sec:4}

The coupling to the SM elementary fields explicitly breaks the global symmetry of the composite sector.
As a consequence the Higgs is only approximately a GB and a potential is generated starting at 1-loop.
Since the breaking is proportional to the mixings the potential is normally dominated by the contributions associated to the top quark.
In the low energy theory with just GBs the contributions to the potential are formally divergent.
More precisely since the lagrangian contains SM interactions one obtains the same quadratic divergences as in the SM
plus additional ones which originate from non-renormalizable interactions, for example Higgs dependent kinetic terms.
We expect that these divergences will be regulated  by the new states in the theory so that the physical
cut-off entering the loops will be the compositeness scale, roughly the mass of the lowest resonances
running in the loop.  This is exactly what happens in QCD for the electro-magnetic splitting of the pions,
see \cite{continoreview}. The same intuition is also correct here.

In the 5D constructions the Higgs effective potential can be calculated at 1-loop and is finite. This property can be traced to the locality
of the theory: since the divergences originate from short distance physics, they cannot be generated as there is
no local gauge invariant operator compatible with the symmetries which contributes to the Higgs mass. The operator which
contributes to the Higgs mass is non-local, being associated to the Wilson line of a gauge field across the fifth dimension.
From this argument it is natural to suspect that the same property will survive in a discretized theory with nearest neighbor interactions
if a sufficient number of sites is included. Interestingly in the model under consideration with a single multiplet of resonances
the 1-loop potential is finite, both for the gauge and the fermionic contributions. The same effect was also found in \cite{panico-wulzer}
in a slightly different setup with an extra layer of resonances. In that paper a general argument explaining the appearance of divergent
terms was also given. Let us mention that exact finiteness is spoiled in the fermionic sector if terms without $LR$ structure
(but compatible with the symmetries) are added to the action (\ref{CHM5}).
The divergence in this case is  logarithmic and can be estimated using as cut-off
the compositeness scale. At any rate exact finiteness of the potential will be always spoiled as we increase the number
of loops so it will always be necessary to treat these computation in an effective field theory sense.

We consider in detail the potential of the CHM$_5$ model described in the previous section.
To leading order the functional form of the potential is fixed by the structure of the symmetry breaking effects \cite{custodian},
\begin{equation}
\label{pot_approx}
V(h) \approx \alpha \, s_h^2  - \beta \, s_h^2 c_h^2,
\end{equation}
where $s_h=\sin h/f$. The coefficients can be expressed as integrals of the correlations functions of the strong sector,
see appendix \ref{sec:B} for all the details. The presence of resonances in our setup renders the integrals finite with the individual terms
saturated by the scale associated to the resonances.

In absence of tuning the natural location of the minimum of the potential is at $\langle h \rangle=f$ or at 0.
In order to achieve successful EWSB ($v<f$) a tuning of the fundamental parameters of the lagrangian must be performed
so that the coefficient of the quadratic term in the expansion of (\ref{pot_approx}) is smaller than its natural size and is
of course negative. In this specific model this requires,
\begin{equation}
\label{yLR}
\frac {\Delta_{t_L}}{m_T}\sim \, \frac {\Delta_{t_R}}{m_{\widetilde T}},
\end{equation}
i.e. similar mixings for left and right chirality. Once the tuning is performed the Higgs mass is given by,
\begin{equation}
\label{vev_mh}
\sin \frac{\langle h \rangle}{f} =  \sqrt{\frac{\beta-\alpha}{2\beta}},\qquad m^2_H = \frac{2}{f^2}\frac{\beta^2-\alpha^2}{\beta}.
\end{equation}

\begin{figure}[ht]
\begin{center}
\subfigure
{\includegraphics[width=0.48\textwidth]{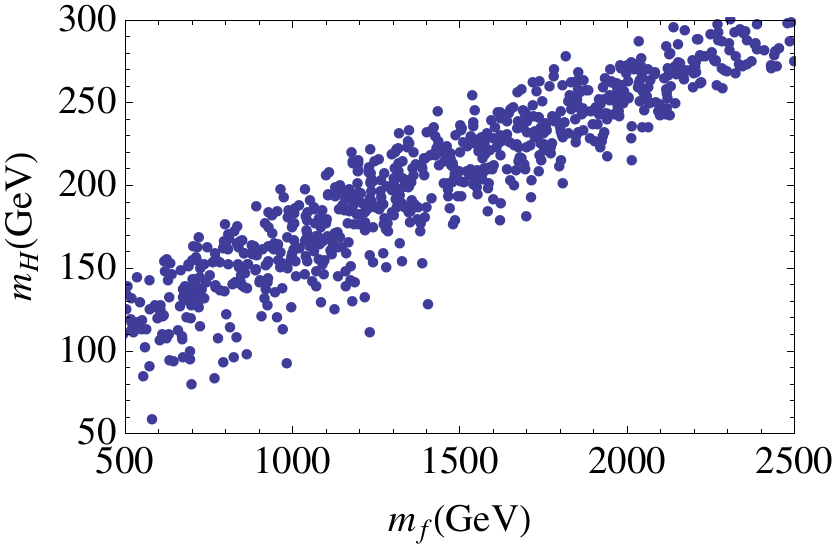}}\quad
\subfigure
{\includegraphics[width=0.48\textwidth]{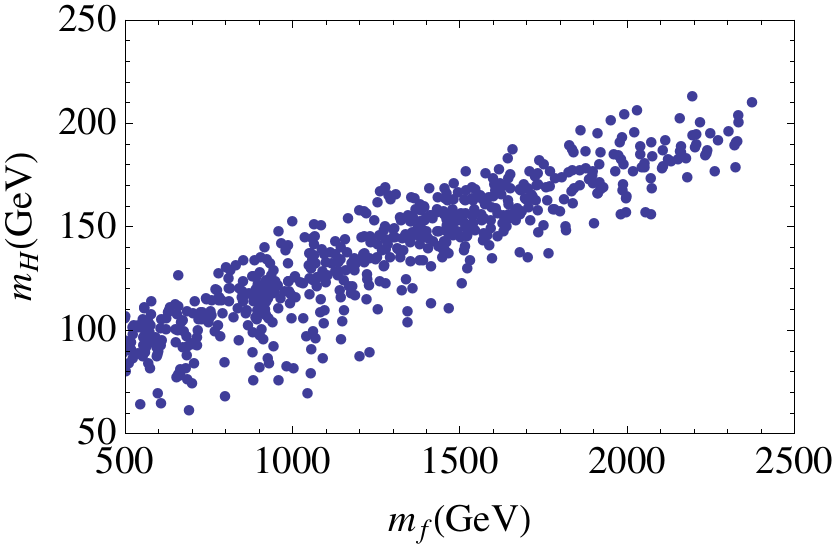}}
\caption{\label{fig:5}\small Masses of the lightest fermionic partners as a function of the Higgs mass for $m_t\in [165,175]$ GeV.
On the left $f=500$ GeV and on the right $f=800$ GeV. The six fermionic parameters are varied between $.5$ and $3$ TeV.
The gauge contribution corresponds to $f_1=f_2=\sqrt{2}f$  and $m_\rho=2.5$ TeV.}
\end{center}
\end{figure}

To determine the Higgs mass as a function of the resonance masses we have performed a scan over the parameters of the model
demanding that the correct electro-weak VEV and top mass are generated.  In figure \ref{fig:5} we present
the results for the unconstrained scan over the six fermionic parameters  $\Delta_{t_L}$, $\Delta_{t_R}$, $m_T$, $m_{\widetilde T}$, $m_{Y_T}$ and $Y_T$
for $f=500$ GeV and $f=800$ GeV. In general we find a strong correlation between Higgs mass and the mass of the lightest resonances, which are typically but not always the ${\bf 2}_{7/6}$ states. In particular if the Higgs is light and the tuning is not large the resonances must be light. This is in agreement with 5D models \cite{custodian,serone}.
As we increase $f$ a Higgs above 115 GeV requires heavy top partners.  In the low mass region we find that the contribution of gauge fields is not necessarily negligible so it is included.

\begin{figure}[t!]
\begin{center}
\subfigure
{\includegraphics[width=0.6\textwidth]{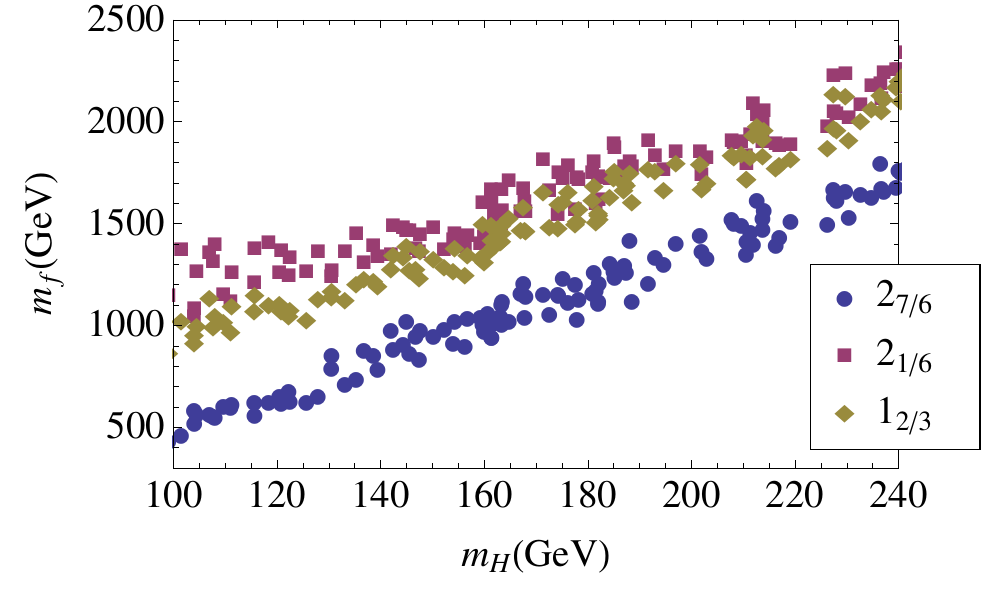}}\quad
\subfigure
{\includegraphics[width=0.6\textwidth]{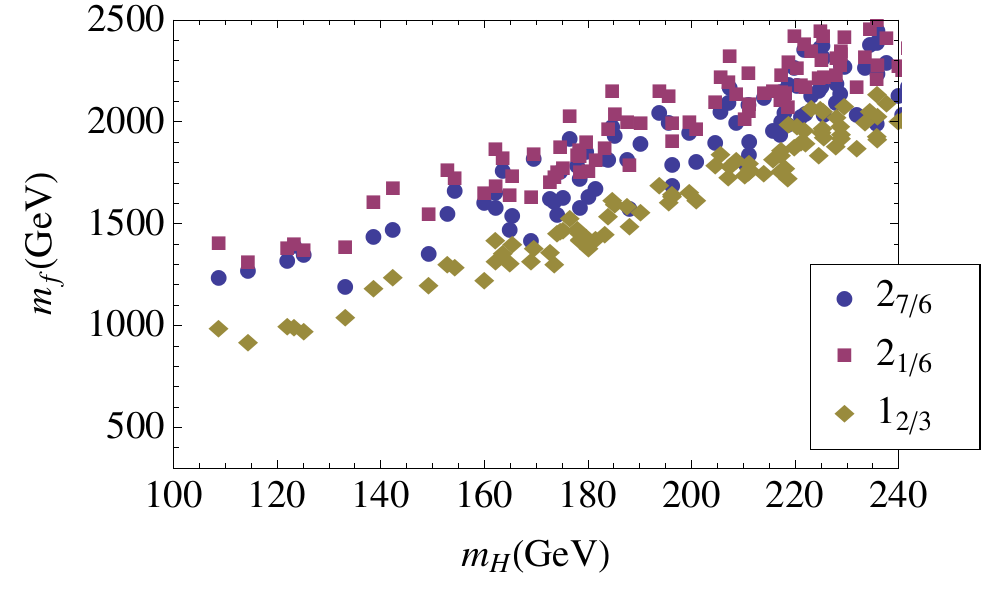}}
\caption{\label{fig:7}\small Masses of lightest fermionic excitations as a function of the Higgs mass for $m_t\in [165,175]$ GeV and $f=500$ GeV.  In the upper plot the mixings are $1.2 \leq \Delta_{t_L}/m_T \leq 1.8$,  $0.7 \leq \Delta_{t_R}/m_{\widetilde T}\leq 1.3$ and the IR parameters  $0.5\le Y_T(\rm{TeV}) \le 3$, $-1.2 Y_T \le  m_{Y_T} \le - 0.8 Y_T$. In the lower plot the mixings are $0.6 \leq \Delta_{t_L}/m_T \leq 0.9$, $0.35 \leq \Delta_{t_R}/m_{\widetilde T}\leq 0.7$  and the IR parameters  $0.5\le Y_T (\rm{TeV})\le  3$, $-0.5\le m_{Y_T}(\rm{TeV}) \le  0.5$.}
\end{center}
\end{figure}

Sharper predictions can be obtained making assumptions on the parameters. In figure \ref{fig:7} we present two scans for different choices of mixings and the IR parameters $Y_T$ and $m_{Y_T}$.
In the first figure with $Y_T\sim -m_{Y_T}$ and large mixings the lightest fermionic partners are the ${\bf 2}_{7/6}$ states.
In the second figure we consider the case $m_{Y_T}\ll Y_T$ and smaller mixings, finding that the singlet  ${\bf 1}_{2/3}$ is the lightest state.

The Higgs potential can be understood from naturalness arguments, see \cite{custodian} and \cite{panico-wulzer} for related discussions.
The low energy lagrangian for the Higgs and SM quarks can be easily extracted from the correlators of the composite sector
in appendix \ref{sec:A} and is given in eq.~(\ref{fermgen}). Its non-linear structure is fixed by the global symmetries.
To lowest order in derivatives we have Yukawa interactions,
\begin{equation}
{\cal L}_{Yuk}= y_t f\, \frac {s_h c_h}{ h} ( \bar q_L H^c t_R+ h.c.)
\end{equation}
and field dependent kinetic terms for the fermions,
\begin{equation}
{\cal L}_{kin}=\frac  {y_{t_L}^2}{2 y_T^2} s_h^2 \, \bar {t}_L  \slashed{D} t_L+ \frac {y_{t_R}^2}{y_{\widetilde{T}}^2} c_h^2 \,
\bar {t}_R \slashed{D} t_R,
\end{equation}
where $y_{T,{\widetilde T}}^{-2}=\widehat{\Pi}_1^{q_L,u_R}(0)$ in (\ref{SE_5y}).

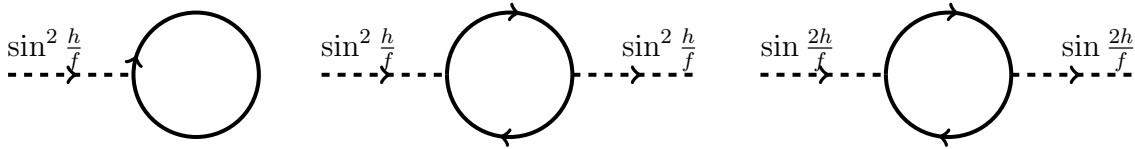
\begin{figure}
\begin{center}
\begin{tikzpicture}[line width=1.5 pt, scale=1.65]
	\draw[scalar] (-1,0)--(0,0);
	\draw[fermion] (1,0) arc (360:0:0.5);
	\node at (-0.7,0.2) {$\sin^2 \frac h f$};
	\draw[scalar] (1.5,0)--(2.5,0);
	\draw[fermion] (2.5,0) arc (180:0:0.5);
	\draw[fermion] (3.5,0) arc (0:-180:0.5);
	\draw[scalar] (3.5,0)--(4.5,0);
	\node at (1.8,0.2) {$\sin^2 \frac h f$};
	\node at (4.2,0.2) {$\sin^2 \frac h f$};
	\draw[scalar] (5,0)--(6,0);
	\draw[fermion] (6,0) arc (180:0:0.5);
	\draw[fermion] (7,0) arc (0:-180:0.5);
	\draw[scalar] (7,0)--(8,0);	
	\node at (5.3,0.2) {$\sin \frac {2h}f$};
	\node at (7.7,0.2) {$\sin \frac {2h}f$};
\end{tikzpicture}
\caption{\label{fig:6}\small Feynman diagrams contributing to the Higgs potential. The first two originate from
Higgs dependent kinetic terms, while the third is obtained from Yukawa interactions.}
\end{center}
\end{figure}

Performing a loop of fermions as in figure \ref{fig:6}, we obtain a contribution to the Higgs potential which is
divergent within the low energy effective theory. We expect the divergence to be cut-off by the compositeness scale
which for the top loops will be the mass of the lightest top partners. From the loop associated to the Yukawa interactions we find,
\begin{equation}
V(h)_{Yuk}\sim N_c \frac {y_t^2}{4\pi^2} m_f^2 f^2\, s_h^2 c_h^2
\label{yukpot}
\end{equation}
where $N_c=3$ is the QCD color factor. From the kinetic terms we obtain to leading order in $y_{t_L,t_R}$,
\begin{equation}
V(h)_{kin}^{(1)} \sim   N_c  \frac {2 y_{t_R}^2- y_{t_L}^2}{32 \pi^2} \frac {m_f^4}{y_T^2}\, s_h^2.
\label{kinpot1}
\end{equation}
There are also sub-leading contributions from the second diagram in figure \ref{fig:6},
\begin{equation}
V(h)_{kin}^{(2)}\sim N_c \frac {y_{t_{L,R}}^4}{16 \pi^2} \frac {m_f^4}{y_T^4}\, s_h^4.
\label{kinpot2}
\end{equation}

To obtain a Higgs VEV $\langle h \rangle< v$ we need to tune the different contributions.
Let us focus on the case $y_{t_L}\sim y_{t_R}$. The leading contributions are the ones
proportional to $y_{t_{L,R}}^2$,  however they tend to cancel for (\ref{yLR}). Due to the different functional dependence of
$V(h)_{Yuk}$ and $V(h)_{kin}^{(2)}$ we can obtain a small Higgs VEV by tuning these terms versus  $V(h)_{kin}^{(1)}$.
The quartic of the potential is then determined by the sub-leading contributions (\ref{yukpot}), (\ref{kinpot1}).
We estimate,
\begin{equation}
m_H\sim  0.3 \,y_t \frac {m_f}{f} v
\end{equation}
in rough agreement with figure \ref{fig:5}.
In particular we see very clearly that, for a given value of $f$, the mass of the Higgs depends linearly on the mass of the top partners.
For a small value of $f$, which is required if the tuning is small, the presence of a light Higgs would then imply
light top partners. Present bounds, with some model dependence, require $m_f > 500$ GeV \cite{D0,CMS}.
Within this model top partners could be soon within the reach of the LHC if the Higgs is light.

One interesting fact is that the contribution of gauge fields, normally sub-leading, is not always negligible in the region
of light fermionic partners. Along the same lines as above we can estimate the contribution of gauge
loops as (see (\ref{explicit-pot-gauge}) for the exact expression),
\begin{equation}
V(h)_{gauge}\sim \frac 9 4 \frac {g_0^2}{16\pi^2}\frac {m_\rho^4}{g_\rho^2}\sin^2 \frac h f,
\end{equation}
where $m_\rho$ is the mass of the spin-1 resonances. Contrary to the fermionic resonances
$m_\rho$ cannot be very low  because of precision electro-weak tests, in particular contributions to the $S$ parameter.
As a consequence the gauge  contribution can become important if the top partners are light, $m_f\ll m_\rho$.
Moreover the gauge contribution is strictly positive so that it might off-set the instability of the potential at the
origin. In general we find that the contribution of gauge loops tends to reduce the number of points with successful
EWSB.

\section{Non-Minimal Interactions}\label{sec:5}

In this section we wish to briefly discuss the effect of adding the non-nearest neighbor interaction
of section \ref{sec:2.3} to the 2-site lagrangian. The role of this term in Higgsless theories \cite{higgsless}
based on the pattern $SO(4)/SO(3)$ was recently considered in \cite{Falkowski:2011ua}
and a related discussion for $SO(5)/SO(4)$ appears in \cite{Contino:2011np}.
The term in eq.~(\ref{f0term}) can be written explicitly as
\begin{equation}
\frac{f_0^2}{2}(D_\mu \Phi)(D^\mu\Phi)^T,
\label{omegaphi}
\end{equation}
where $\Phi=\Phi_2\Omega_1^T$ as in eq.~(\ref{GBSO5}). This contributes to the kinetic term of the GBs but not to the mass of the
composite resonances in eq.~(\ref{masses1}). Importantly it changes the relation between the gauge coupling of the resonances
and their coupling to GBs. For $SO(4)$ resonances, in the limit $f_2\to \infty$, one finds\footnote{We define $g_{\rho\pi\pi}$ from the interaction
$g_{\rho\pi\pi} f^{\widehat{a}\widehat{b}c} \partial_{\mu} \pi^{\widehat{a}} \pi^{\widehat{b}}\rho_\mu^c$.
Equivalent formulas to eqs.~(\ref{grhopipi}), (\ref{KRSF}) were recently obtained  in \cite{Contino:2011np} in the CCWZ language.
Note that there are corrections to these equations for finite $f_2$, see appendix \ref{appwidth}.},
\begin{equation}
g_{\rho\pi\pi}=  \frac {f^2-f_0^2}{2 f^2}\, g_\rho
\label{grhopipi}
\end{equation}
while for coset resonances there is no coupling to two GBs by $SO(4)$ symmetry.
From this expression, using eq.~(\ref{masses1}), one obtains
\begin{equation}
m_\rho^2= 2  \frac {f^2}{f^2-f_0^2}\,g_{\rho\pi\pi}^2\,f^2
\label{KRSF}
\end{equation}
which is the analog of the KRSF relation in QCD. The modified coupling of the $\rho$ to the Higgs can
have important phenomenological effects. First of all it modifies the decay of the resonances
into $W^+_LW^-_L$ and $Z_L h$ which together with third generation quarks are the main decay channel. Since the decay
goes like $g_{\rho\pi\pi}^2$ this term can easily generate an order one change in the decay, see appendix \ref{appwidth} for more
details.

Importantly the term (\ref{omegaphi}) also contributes to the $S-$parameter.
This can be readily computed in terms of the two-point functions of the currents of the strong sector \cite{Contino},
\begin{equation}
S= 4\pi \Pi_1'(0)  \frac {v^2} {f^2}.
\end{equation}
Including the term proportional to $f_0^2$ amounts to adding a constant in $\Pi_1$ so that the dependence of $S$ on $f_0$ (for fixed resonance masses)
appears through the redefinition of $f$ in eq.~(\ref{fhls}). The result can be written as,
\begin{equation}
S= 4 \pi v^2 \left(\frac 1 {m_\rho^2}+ \frac 1 {m_{a_1}^2}\right)  \displaystyle{\frac {f^2-f_0^2}{f^2}}.
\label{S}
\end{equation}
We note however that the $S$ parameter could be suppressed if $f_0^2$ is sizable and positive.\footnote{We should mention that in
QCD the sign of the analogous term is negative which would correspond to an enhancement of the $S-$parameter.
It would be interesting to know whether the sign can be robustly fixed in a strongly coupled theory not QCD-like.}
In particular $S$ would be zero if $f_0=f$.  In this limit, which can be realized taking $f_2=0$ keeping the spectrum finite,
GBs and resonances are decoupled from each other, the latter becoming degenerate $SO(5)$ multiplets.
A similar effect was noticed in ref.~\cite{vanishingS}, in the context of Higgsless theories.
Indeed our setup with symmetry structure $SO(4)/SO(3)$ can be used as 2-site model for Higgsless and
in this case the contribution to $S$ would be given by eq.~(\ref{S}) with $f=v$.
From the point of view of the effective theory $S$ could even be negative. This can be achieved if $f_0> f$ which
corresponds to having the mass of the coset resonances lighter than $SO(4)$ resonances.
While these extreme values of $f_0$ might be questionable, since the resonances not even partially
unitarize the scattering of GBs, for phenomenological purposes even a mild suppression can be important.

\section{Conclusions}
\label{sec:6}

In this note we have presented a simplified 4D description of Composite Higgs Models with partial compositeness.
Our formalism allows to introduce an arbitrary tower of resonances, bosonic and fermionic, in a theory where the Higgs is a GB.
As a particular case one can reproduce the physics of higher dimensional  models.

For phenomenological applications we focused on the minimal model with only the lowest lying resonances
which provides a self-contained framework reproducing all the relevant features of Composite Higgs Models at the LHC.
For example our theoretical setup already allows to compute the Higgs potential with a minimal number of degrees
of freedom. We believe that this is the most natural and economic setup, being dictated by the symmetries of theory.
As an example we considered in detail the CHM$_5$ scenario of \cite{custodian} with results comparable to the 5D scenario
for similar choices of parameters.

While our construction was inspired by 5D models one can consider the truncation of the theory
to the lowest resonances on its own.  Based on the symmetries of the theory  more terms can be
added to the lagrangian which do not follow from the discretization of the 5D theory. In particular
we have shown that, as for the $\rho$ in QCD, these terms modify the coupling of the resonances to
the GBs and could be of phenomenological relevance changing their decays and
possibly reducing the $S-$parameter.

Our work is the starting point for a study of the model at LHC.  The simplified lagrangian allows to capture
all the effects of compositeness and is simple enough to be implemented on an event generator.
We hope to return to this in future work.

\subsection*{Acknowledgments}
We would like to thank Roberto Contino, Leandro da Rold and Andrea Wulzer for useful discussions.
AT would like to thank CERN for partial support and hospitality.

\appendix

\section{Self-Energies}
\label{sec:A}

In this appendix we provide explicit expressions for the two-point functions of the strong sector
in the 2-site picture. We follow closely ref.~\cite{Contino} where this technique was introduced.
The great advantage of this approach is that the relevant informations can be extracted from the
correlators of the composite sector computed for zero Higgs VEV (since the composite sector does not
depend on it). We analyze the gauge and fermionic sectors of the theory presenting the formulas
for CHM$_5$ and CHM$_{10}$ models of \cite{custodian}.

\subsection{Spin-1}
In the background of a constant GB Higgs the most general quadratic lagrangian for the vector sources
invariant under $SO(5)\times U(1)_X$ takes the form,
\begin{equation}
\begin{split}
{\cal L}_{\rm eff}
=& \frac{1}{2}P^T_{\mu\nu}\Big[ \widehat\Pi^{X}_0(p^2)\, X^{\mu} X^{\nu}
+\widehat\Pi_0(p^2)\, {\rm Tr}\big[A^{\mu} A^{\nu}\big]
+\widehat\Pi_1(p^2)\, \Phi A^{\mu} A^{\nu} \Phi^T\Big],
\label{eq_eff}
\end{split}
\end{equation}
where $P^T_{\mu\nu}= \eta_{\mu\nu}-p_{\mu}p_{\nu}/p^2$, $A^A,X \in Adj[SO(5)\times U(1)_X]$ and $\Phi$ is defined in eq.~(\ref{GBSO5}). To determine the form factors we need to express the lagrangian (\ref{gaugeminimal})
as a function of $A^{a,\widehat{a}}$ and $X$ by integrating out the fields $\rho^\mu$ according to their
equations of motion. The result, computed for $h=0$, is:
\begin{equation}
{\cal L}_{\rm eff} = \frac{1}{2}P^T_{\mu\nu}\left[ \Pi_a(p^2)\ A^{\mu}_a A^{\nu}_a + \Pi_{\widehat{a}}(p^2) A^{\mu}_{\widehat{a}} A^{\nu}_{\widehat{a}} +\Pi_X(p^2)X^{\mu}X^{\nu}  \right]
\label{eq_eff5}
\end{equation}
where\footnote{The formulas here and below are well defined for euclidean momenta and the appropriate analytic continuation
to Lorentzian signature is understood.},
\begin{equation}
\begin{aligned}
\label{Pi_aa}
\Pi_a(p^2)&= \frac{m_{\rho}^2 p^2}{g_\rho^2 \left(p^2-m_{\rho}^2\right)}, \\
\Pi_{\widehat{a}}(p^2)&= \frac{m_{\rho}^2 \left(p^2-(m_{a_1}^2-m_{\rho}^2)\right)}{g_\rho^2 \left(p^2-m_{a_1}^2\right)}, \\
\Pi_X(p^2)&= \frac{m_{\rho_X}^2 p^2}{g_{\rho_X}^2 \left(p^2-m_{\rho_X}^2\right)}.
\end{aligned}
\end{equation}
The form factors in (\ref{eq_eff}) are identified as,
\begin{equation}
\label{Pi5D-4D}
\widehat\Pi_0(p^2)= \Pi_a(p^2),\quad \widehat\Pi^X_0(p^2)= \Pi_X(p^2),\quad \widehat \Pi_1(p^2) = 2\left[\Pi_{\widehat{a}}(p^2)-\Pi_a(p^2)\right].
\end{equation}
To this we have to add the kinetic terms for the elementary gauge fields.\footnote{These terms are not necessary in the 5D realization of the model since a kinetic term for
the zero modes will be generated by the volume of the extra-dimension.}
Therefore we define,
\begin{eqnarray}
\label{Pi_0}
\Pi_0(p^2)&=& -\frac{p^2}{g_0^2}+ \widehat \Pi_0(p^2)= -\frac{p^2}{g_0^2} + \frac{m_{\rho}^2 p^2}{g_\rho^2 \left(p^2-m_{\rho}^2\right)}\,, \nonumber \\
\label{Pi_Y}
\Pi_Y(p^2)&=& -\frac{p^2}{g_{0Y}^2}+\widehat\Pi_0(p^2)+ \widehat\Pi_0^X(p^2)= -\frac{p^2}{g_{0Y}^2} + \frac{m_{\rho}^2 p^2}{g_\rho^2 \left(p^2-m_{\rho}^2\right)} + \frac{m_{\rho_X}^2 p^2}{g_{\rho_X}^2 \left(p^2-m_{\rho_X}^2\right)}\,, \nonumber \\
\label{Pi_1hat}
\Pi_1(p^2)&=& \widehat\Pi_1(p^2) = -\frac{2 m_{\rho}^4 \left(m_{\rho}^2-m_{a_1}^2\right)}{g_\rho^2 \left(p^2-m_{a_1}^2\right) \left(p^2-m_{\rho}^2\right)}.
\end{eqnarray}
Setting to zero the non dynamical fields we obtain the coupling of the Higgs to the SM gauge fields,
\begin{eqnarray}
\label{low-gauge}
{\cal L}^{gauge}_{\rm eff} &=& \frac{1}{2} P^T_{\mu\nu} \bigg[
 \left( \Pi_0(p^2) + \frac{\sin^2(h/f)}{4} \Pi_1(p^2) \right) A^\mu_{aL} A^\nu_{aL} \nonumber \\
 &+& \left( \Pi_Y(p^2) + \frac{\sin^2(h/f)}{4}\Pi_1(p^2) \right) Y^\mu Y^\nu + 2 \sin^2(h/f)\, \Pi_1(p^2)  \;
     \widehat H^\dagger T^{a}_L Y \widehat H\, A^\mu_{aL} Y^\nu \bigg], \nonumber \\
\end{eqnarray}
where $A^{aL}_\mu, Y_\mu$ are $SU(2)_L\times U(1)_Y$ gauge fields and $\widehat H=\displaystyle\frac{1}{h}\left(\begin{array}{c}h^2-ih^1\\ h^4-ih^3\end{array} \right)$. Expanding in $p^2$ we extract the $W$ and $Z$ masses:
\begin{equation}
\begin{aligned}
m^2_W&= - \frac{1}{4}\frac{\Pi_1(0)}{\Pi_0'(0)}\sin^2 \frac{\langle h \rangle}{f}=\frac {g^2}{4}f^2\, \sin^2 \frac{\langle h \rangle}{f} , \\
m^2_Z&= - \frac{1}{4}\left[\frac{\Pi_1(0)}{\Pi_0'(0)}+\frac{\Pi_1(0)}{\Pi_Y'(0)}\right]\sin^2 \frac{\langle h \rangle}{f}=\frac {g^2+g'^2 }{4}f^2\, \sin^2 \frac{\langle h \rangle}{f}.
\end{aligned}
\end{equation}
from which the definition of $v$ in (\ref{VEV}) follows.

\subsection{Spin-1/2}

The effective action for SM fermions for CHM$_5$ and CHM$_{10}$  takes the form  \cite{custodian},
\begin{equation}
\begin{aligned}
{\cal L}_{\rm eff} &=
\bar q_L \slashed{p} \left[ \Pi_0^q(p^2)
 + \frac{s^2_h}{2} \left( \Pi_1^{q1}(p^2)\, \widehat H^c \widehat H^{c\dagger}
 +  \Pi_1^{q2}(p^2)\, \widehat H \widehat H^\dagger \right) \right] q_L \\
& +\bar u_R \slashed{p} \left( \Pi_0^u(p^2) + \frac{s^2_h}{2}\, \Pi_1^u (p^2)\right) u_R
 +\bar d_R \slashed{p} \left( \Pi_0^d (p^2)+ \frac{s^2_h}{2}\, \Pi_1^d (p^2)\right) d_R  \\
&+ \frac{s_hc_h}{\sqrt{2}} M_1^u (p^2)\,\bar q_L \widehat H^c u_R
 + \frac{s_hc_h}{\sqrt{2}} M_1^d (p^2)\,\bar q_L \widehat H d_R + h.c. \,  .
\end{aligned}
\label{fermgen}
\end{equation}
We find that the form factors can be expressed in both models in terms of the following functions,
\begin{equation}
\label{bido-sing}
\begin{aligned}
\widehat{\Pi}[m_1,m_2, m_3]&= \frac{\left(m_2^2+m_3^2-p^2\right)\Delta ^2 }{p^4 - p^2 (m_1^2+m_2^2+m_3^2) +m_1^2m_2^2},\\
\widehat{M}[m_1, m_2, m_3]&=- \frac{m_1 m_2 m_3\,\Delta ^2}{p^4 - p^2 (m_1^2+m_2^2+m_3^2) +m_1^2m_2^2}.
\end{aligned}
\end{equation}

\subsubsection{CHM$_5$}

To determine the form factors in (\ref{fermgen}) we write the $SO(5)\times U(1)_X$ invariant action
for the sources $\Psi_{q_L}$ and $\Psi_{u_R}$ (neglecting the bottom sector),
\begin{eqnarray}
\label{CHM5-eff}
{\cal L}^{{\rm CHM_5}}_{{\rm eff}} &=&\ \!
 \bar{\Psi}_{q_L}^i \slashed{p} \left( \delta^{ij} \widehat{\Pi}^{q_L}_0(p^2)
  + \Phi^i \Phi^j \widehat{\Pi}_1^{q_L}(p^2) \right)\! \Psi_{q_L}^j
  + \! \bar{\Psi}_{u_R}^i \slashed{p} \left( \delta^{ij} \widehat{\Pi}_0^{u_R}(p^2)
  + \Phi^i \Phi^j \widehat\Pi_1^{u_R}(p^2) \right)\! \Psi_{u_R}^j \nonumber \\
 &+& \bar{\Psi}_{q_L}^i \!\left( \delta^{ij} \widehat{M}_0^{u}(p^2)
   +\Phi^i \Phi^j \widehat{M}_1^{u}(p^2) \right)\! \Psi_{u_R}^j
 + h.c.
\end{eqnarray}
The form factors can be obtained integrating out the composite fermions in (\ref{fermions}).
In terms of the building blocks (\ref{bido-sing}) one derives,
\begin{eqnarray}
\label{SE_5y}
\widehat{\Pi}^{q_L}_0&=&\widehat{\Pi}[m_T, m_{\widetilde{T}}, m_{Y_T}]\,,~~~~~~~\widehat{\Pi}^{q_L}_1=\widehat{\Pi}[m_T, m_{\widetilde{T}}, m_{Y_T}+Y_T]-\widehat{\Pi}[m_T, m_{\widetilde{T}}, m_{Y_T}], \nonumber \\
\widehat{\Pi}^{u_R}_0&=&\widehat{\Pi}[m_{\widetilde{T}}, m_T , m_{Y_T}]\,,~~~~~~~\widehat{\Pi}^{u_R}_1=\widehat{\Pi}[m_{\widetilde{T}}, m_T , m_{Y_T}+Y_T]-\widehat{\Pi}[ m_{\widetilde{T}}, m_T , m_{Y_T}], \nonumber \\
\widehat{M}^{u}_0&=& \widehat{M}[m_T, m_{\widetilde{T}}, m_{Y_T}]\,,~~~~~~\widehat{M}^{u}_1=\widehat{\Pi}[m_T, m_{\widetilde{T}}, m_{Y_T}+Y_T]-\widehat{\Pi}[m_T, m_{\widetilde{T}}, m_{Y_T}].\nonumber\\
\end{eqnarray}
The matching with eq.~(\ref{fermgen}) is obtained for:
\begin{equation} \label{CHM5-Piq}
\begin{split}
\Pi_0^q &= \frac{1}{y_{t_L}^2}+ \widehat\Pi_0^{q_L} \, , \\
\Pi_0^u &=\frac{1}{y_{t_R}^2}+ \widehat\Pi_0^{u_R} + \widehat\Pi_1^{u_R} \, , \\
\end{split} \qquad
\begin{split}
&\Pi_1^{q_1}= \widehat\Pi_1^{q_L}\, , \\
&\Pi_1^{u}= -2\, \widehat\Pi_1^{u_R}\
\end{split} \qquad
\begin{split}
M_1^u &= \widehat M_1^{u}\, , \\[0.05cm]
\end{split}
\end{equation}
where we have added the kinetic terms for the elementary  fields. In our CHM$_5$, $\Pi_1^{q_2}$ in (\ref{fermgen}) is negligible
due to the choice $\Delta_{b_L} \ll \Delta_{t_L}$.

From eq.~(\ref{fermgen}) we extract the top mass,
\begin{eqnarray}
\label{massatop}
m_{t}&\approx& \frac{s_h c_h}{\sqrt{2}}\frac{M^{u}_1(0)}{\sqrt{\Pi^{q}_0(0) \Pi^{u}_0(0)}} \nonumber \\
&\approx & \frac{v}{\sqrt{2}}\cdot  \frac{1}{ \sqrt{\left(1+\frac{\left(m_{\widetilde T}^2+m_{Y_T}^2\right) \Delta_{t_L}^2}{m_T^2 m_{\widetilde T}^2}\right) \left(1+\frac{\left(m_T^2+(m_{Y_T}+Y_T)^2\right) \Delta_{t_R}^2}{m_T^2 m_{\widetilde T}^2}\right)}}\cdot \frac{\Delta_{t_L}}{m_T} \frac{\Delta_{t_R}}{m_{\widetilde{T}}} \frac  {Y_T}f\nonumber \\
\end{eqnarray}
used for the numerical computations in section \ref{sec:4}.

\subsubsection{CHM$_{10}$}

Here we briefly present the relevant formulas for the CHM$_{10}$ model of \cite{custodian}.
We refer to this paper for all the details. In this model the SM quark are embedded in three distinct {\bf 10} reps with $X=2/3$.
Under $SU(2)_L\times SU(2)_R$ the representation decomposes as ${\bf 10}= (\bf{2},\bf{2})\oplus ({\bf 3},{\bf 1})\oplus ({\bf 1},{\bf 3})$.
Differently from the CHM$_5$ case, SM singlet $t_R$ and $b_R$ couple to a $({\bf 1},{\bf 3})$ with $T^3_R=0,-1$ respectively.

The 2-site CHM$_{10}$ lagrangian reads,
\begin{equation}
\label{CHM10}
\begin{aligned}
{\cal L}^{\rm CHM_{10}} &= {\cal L}^{el}_{fermions}\\
&+ \Delta {\rm Tr}\left[\bar{q}_L \Omega_1 \Psi_T \right] + \Delta {\rm Tr}\left[\bar{t}_R \Omega_1 \Psi_{\widetilde T} \right] +h.c.\\
&+ {\rm Tr}\left[\bar{\Psi}_T \left(i\slashed{D}^{\rho} - m_T \right)\Psi_T \right] + {\rm Tr}\left[\bar{\Psi}_{\widetilde T} \left(i\slashed{D}^{\rho} - m_{\widetilde T} \right)\Psi_{\widetilde T} \right] \\
&- Y_T \Phi_2 \bar{\Psi}_{T,L} \Psi_{\widetilde T, R}\Phi_2^T - m_{Y_T}{\rm Tr}\left[ \bar{\Psi}_{T,L} \Psi_{\widetilde T, R} \right]+h.c.
\end{aligned}
\end{equation}
The effective lagrangian for the chiral sources $\Psi_{q_L}$ and $\Psi_{u_R}$ in
the {\bf 10} rep, in the background of a constant GB field, is
\begin{equation}
\label{CHM10eff}
\begin{split}
{\cal L}^{\rm CHM_{10}}_{\rm eff}
 = &\sum_{r=q_L,u_R}\Big[{\rm Tr}\big(\bar \Psi_r \slashed{p} \, \widehat\Pi^{r}_0(p^2) \Psi_{r}\big)
 +\Phi\, \bar\Psi_{r}\slashed{p} \, \widehat\Pi_{1}^r(p^2) \Psi_{r}\Phi^T \Big] \\
  &+\Big[{\rm Tr}\big(\bar \Psi_{q_L} \widehat M_0^{u}(p^2) \Psi_{u_R}\big)
 +\Phi\, \bar \Psi_{q_L} \widehat M_1^{u}(p^2) \Psi_{u_R}\Phi^T \Big] + h.c. \, .
\end{split}
\end{equation}
Also in this case we can write the form factors in terms of (\ref{bido-sing}),
\begin{eqnarray}
\label{CHM10PI}
\widehat{\Pi}^{q_L}_0&=&\widehat{\Pi}[m_T, m_{\widetilde{T}}, m_{Y_T}]\,,~~~~~~~\widehat{\Pi}^{q_L}_1=2 \widehat{\Pi}[m_T, m_{\widetilde{T}}, m_{Y_T}+Y_T/2]-2\widehat{\Pi}[m_T, m_{\widetilde{T}}, m_{Y_T}] ,\nonumber \\
\widehat{\Pi}^{u_R}_0&=&\widehat{\Pi}[m_{\widetilde{T}}, m_T , m_{Y_T}]\,,~~~~~~~\widehat{\Pi}^{u_R}_1=2 \widehat{\Pi}[m_{\widetilde{T}}, m_T , m_{Y_T}+Y_T/2]- 2\widehat{\Pi}[ m_{\widetilde{T}}, m_T , m_{Y_T}], \nonumber \\
\widehat{M}^{u}_0&=& \widehat{M}[m_T, m_{\widetilde{T}}, m_{Y_T}]\,,~~~~~~\widehat{M}^{u}_1=2 \widehat{\Pi}[m_T, m_{\widetilde{T}}, m_{Y_T}+Y_T/2]-2 \widehat{\Pi}[m_T, m_{\widetilde{T}}, m_{Y_T}].\nonumber\\
\end{eqnarray}
To match with eq.~(\ref{fermgen}) we define:
\begin{equation}
\label{CHM10-Piq}
\begin{split}
\Pi_0^q &= \frac{1}{y_{t_L}^2}+ \widehat\Pi^{q_L}_0+\frac{1}{2}\, \widehat\Pi^{q_L}_1 \, , \\
\Pi_0^u &= \frac{1}{y_{t_R}^2}+ \widehat\Pi^{u_R}_0 \, , \\
\end{split} \qquad
\begin{split}
\Pi_1^{q_1}&= -\frac{1}{2}\, \widehat\Pi^{q_L}_1 \, , \\
\Pi_1^{q_2}&= -\widehat\Pi^{q_L}_1 \, , \\
\Pi_1^{u}  &= \frac{1}{2}\, \widehat\Pi^{u_R}_1 \, ,
\end{split} \qquad
\begin{split}
M_1^u &= \frac{1}{2\sqrt{2}}\, \widehat M^{u}_1  \, , \\[0.05cm]
\end{split}
\end{equation}
which include kinetic terms of elementary fields. Differently from CHM$_5$, now $\Pi_1^{q_2}$ is not negligible.

\section{Effective Potential}
\label{sec:B}

In this appendix we provide the details of the computation of the effective potential.
This can be expressed in terms of integrals of the two-point functions of the strong sector listed
in appendix \ref{sec:A}. In particular from the asymptotic behavior of the form factors one
can easily check the convergence of the potential.

\subsection{Gauge Contribution}

We start considering the contribution to the Higgs potential induced by $SU(2)_L$ gauge fields.
Integrating over the gauge fields in (\ref{low-gauge}) one has,
\begin{equation}
\label{Coleman-gauge}
V(h)_{gauge}= \frac{9}{2}\int \frac{d^4 p}{(2\pi)^4}\ln \left[1 +
\frac{1}{4}\frac{\Pi_1(p^2)}{\Pi_0(p^2)}\sin^2 \frac{h}{f} \right],
\end{equation}
where $\Pi_0(p^2)$ and $\Pi_1(p^2)$ are defined in eq.~(\ref{Pi_0}).
The integral is UV convergent because $\Pi_1/\Pi_0$ scales as
\begin{equation}
\label{convergenze-gauge}
\frac{\Pi_1(p^2)}{\Pi_0(p^2)} = -\frac{2 g_0^2 m_{\rho}^4 \left(m_{a_1}^2-m_{\rho}^2\right)}{g_\rho^2 p^2 \left[p^2- m_{\rho}^2\left(1+g_0^2/g_\rho^2\right)\right] \left(p^2-m_{a_1}^2\right)} \stackrel{p^2 \rightarrow \infty}{\longrightarrow} \frac{1}{p^6}.
\end{equation}
To leading order
\begin{eqnarray}
\label{explicit-pot-gauge}
V(h)_{gauge}&\approx&  \int \frac{d^4p}{(2\pi)^4} \frac{9}{8}\frac{\Pi_1}{\Pi_0}\sin^2 \frac h f\nonumber \\
&=& \frac{9}{4}  \frac{1}{16\pi^2} \frac{g_0^2}{g_\rho^2} \frac{m_{\rho}^4 \left(m_{a_1}^2-m_{\rho}^2\right)}{m_{a_1}^2-m_{\rho}^2(1+g_0^2/g_\rho^2)}\ln \left[\frac{m_{a_1}^2}{m_{\rho}^2(1+g_0^2/g_\rho^2)} \right]\sin^2 \frac h f.
\end{eqnarray}
The curvature of the potential at the origin is positive, a general feature of gauge interactions which tend to preserve the symmetry.

For $m_{a_1} \rightarrow \infty$ (corresponding to $f_2\to \infty$ in our setup) the potential becomes logarithmically divergent,
showing that the coset resonances are crucial for the finiteness of the potential.
The finiteness of the potential is spoiled once non-local terms are added. In the language of the
self-energies this corresponds to adding a constant to $\Pi_1$.
One easily sees that in this case the potential is quadratically divergent.

\subsection{Fermionic Contribution}

The fermionic contibution to the Higgs effective potential is \cite{Contino}:
\begin{equation}
\label{pot_fer}
V(h)_{fermions}= -2N_c \int\frac{ d^4p}{(2\pi)^4}\left[\ln \Pi_{b_L} + \ln \left(p^2 \Pi_{t_L}\Pi_{t_R} - \Pi^2_{t_Lt_R} \right)\right].
\end{equation}
From (\ref{fermgen}) we find,
\begin{equation}
\begin{aligned}
\Pi_{t_L}&= \Pi^{q}_0 + \frac{\sin^2 (h/f)}{2} \Pi_1^{q_1}\\
\Pi_{b_L}&= \Pi^{q}_0 + \frac{\sin^2 (h/f)}{2} \Pi_1^{q_2}\\
\Pi_{t_R}&= \Pi_{0}^{u} +\frac{\sin^2 (h/f)}{2} \Pi_1^{u}\\
\Pi_{t_Lt_R} &= \frac{\sin (h/f)\cos (h/f)}{\sqrt{2}}M^{u}_1.
\end{aligned}
\end{equation}
Let us focus on the CHM$_5$. Due to the choice of representations in this model, for $\Delta_{b_L}\ll \Delta_{t_L}$ the contribution to the potential of $\Pi_{b_L}$ is
negligible. Expanding (\ref{pot_fer}), we can identify the coefficients $\alpha$ and $\beta$ of eq.~(\ref{pot_approx}),
\begin{equation}
\label{pot_param}
\alpha = -2N_c \int \frac{d^4p}{(2\pi)^4} \left[ \frac{1}{2}\frac{\Pi_1^{q_1}}{\Pi_0^{q}} + \frac{1}{2}\frac{\Pi_1^{u}}{\Pi_0^{u}}\right],
\end{equation}
\begin{equation}
\beta = -N_c \int \frac {d^4p}{(2\pi)^4} \left[\frac{(M_1^{u})^2}{p^2 \Pi_0^{q} \Pi_0^{u}}\right].
\end{equation}

Using the expressions listed in (\ref{SE_5y}, \ref{CHM5-Piq}) it easy to check that the integrals are finite as each term
goes at most as $1/p^6$ at large momenta. For example,
\begin{equation}
\frac{\Pi_1^{q}}{\Pi_0^{q}}= - \frac {Y_T(2 m_{Y_T}+Y_T) m_T^2 \Delta_{t_L}^2}{p^6} +{\cal O}\left(\frac 1 {p^8}\right).
\end{equation}
For $Y_T\to \infty$ the integral becomes logarithmically divergent.
This is similar to the potential generated by gauge loops when $f_2\to \infty$
since we do not have complete $SO(5)$ multiplets in this limit. This shows that a complete $G$ multiplet of resonances is the
minimal number of degrees of freedom which allows the leading contribution to the potential to be finite.
The finiteness of the potential is also spoiled if the most general lagrangian without $LR$ structure is written on the
last site in eq.~(\ref{CHM5}), again with a logarithmic divergence.

The most important feature of this scenario is that  the potential is tunable because we have
contributions with different functional dependence, on $\sin^2 (h/f)$ and $\sin^2 (h/f)\cos^2 (h/f)$.
Successful EWSB can be obtained for $\Delta_{t_L}\sim \Delta_{t_R}$ as the two contributions
to $\alpha$ in (\ref{pot_param}) tend to cancel each other leading to $\alpha$ smaller than its natural size.
One can also see that the leading contributions to $\alpha$ vanish for $Y_T\to -2 m_{Y_T}$.

CHM$_{10}$ works in a similar way, in particular also in this case the potential is finite.

\section{Interaction Vertices $\rho \pi\pi$}
\label{appwidth}

Here we present the general formulas for trilinear vertices among composite spin-1
resonances and GBs. We focus on the $SO(4)$ resonances as the coset resonances
have suppressed trilinear couplings to GBs (in absence of mixing to the elementary fields they are zero by $SO(4)$ invariance, and only a small coupling is induced after EWSB).
We will include non-minimal interactions (\ref{omegaphi}) which, if sizable, importantly modifies the decay widths.
In the general case the GBs are conveniently parametrized as,
\begin{equation}
\Omega_1 = e^{\frac {i} f \frac {f_2^2}{f_1^2+f_2^2} \Pi}\,,~~~~~~~~~~~~~~~~~~\Phi_2=\phi_0 e^{-\frac {i} f \frac {f_1^2}{f_1^2+f_2^2} \Pi},
\end{equation}
where $f$ is the low energy decay constant (\ref{fhls}).
Expanding the lagrangian (\ref{gaugeminimal}) + (\ref{omegaphi}) around $\Pi=0$ we obtain,
\begin{equation}
\label{cubic2}
\begin{aligned}
{\cal L}_{cubic}&=-\frac{f_0^2}{f^2} g_0\partial_{\mu}\pi^{\widehat{a}}\pi^{\widehat{b}} A_{\mu}^{c'}  f^{\widehat{a} \widehat{b}c'} -\frac{(f^2-f_0^2)^2}{2 f^2 f_1^2 }g_0\partial_{\mu}\pi^{\widehat{a}}\pi^{\widehat{b}} A_{\mu}^{c'}  f^{\widehat{a} \widehat{b}c'}\\
&-\frac{f^2-f_0^2}{2f^2} g_\rho \partial_{\mu}\pi^{\widehat{a}}\pi^{\widehat{b}} \rho_{\mu}^{c}  f^{\widehat{a} \widehat{b}c}
-\frac{(f^2-f_0^2)^2}{2f^2 f_2^2} g_\rho \partial_{\mu}\pi^{\widehat{a}}\pi^{\widehat{b}} \rho_{\mu}^{c}  f^{\widehat{a} \widehat{b}c}.
\end{aligned}
\end{equation}
Here the elementary fields are $A_{\mu}=A_{\mu}^{c'}T^{c'}$, where $T^{c'}=\{T^a_L,Y\}$ and composite ones $\rho_{\mu}=\rho_{\mu}^c T^c$ with $T^c=\{T^a_L,T^a_R,X\}$. We have also traded $h^{\widehat{a}}$ for $\pi^{\widehat{a}}$.
We derive that, in absence of couplings to the elementary fields, the $\rho\pi\pi$ coupling is ,
\begin{equation}
\label{rhopipif2comp}
g_{\rho\pi\pi}= \frac{f^2-f_0^2}{2f^2} \left(1+\frac{f^2-f_0^2}{f_2^2} \right)g_\rho.
\end{equation}
Eq.~(\ref{rhopipif2comp}) shows that in the limit $f_2=0$ the heavy resonances and GBs are decoupled,
while for $f_2\rightarrow\infty$ we recover (\ref{grhopipi}). From this we also obtain
\begin{equation}
m_\rho^2= \frac {f_2^6}{(f_2^4-(f^2-f_0^2)^2)(f_2^2+f^2-f_0^2)}\times  \frac {2 f^2}{f^2-f_0^2}\,g_{\rho\pi\pi}^2\,f^2
\end{equation}
which generalizes (\ref{KRSF}) to finite $f_2$.

To evaluate effects due to the mixing with elementary states, let us consider the $SU(2)_L$ fields. In the mass basis (before EWSB), we have\footnote{Some commutation rules are $[T^{\widehat{a}},T^{\widehat{b}}]=\frac{i}{2}\epsilon^{abc}(T^c_L+T^c_R)$ and $[T^{\widehat{a}},T^{\widehat{4}}]=\frac{i}{2}(T^a_L-T^a_R)$.}:
\begin{equation}
\label{rhopipi2}
\left\{g \cot\theta_L \left[ \frac{f^2-f_0^2}{2f^2}\left(1+\frac{f^2-f_0^2}{f_2^2}\right)\right]-g\tan\theta_L\left[\frac{f_0^2}{f^2}+\frac{(f^2-f_0^2)^2}{2f_1^2 f^2}\right] \right\}\ \partial^{\mu}\pi^{\widehat{a}}\pi^{\widehat{b}}\rho_{\mu}^{c_L} f^{\widehat{a}\widehat{b}c_L},
\end{equation}
where $g$ is defined in (\ref{couplingL}) and $\theta_L$ in (\ref{massL}). To compare with the results of \cite{contino-sundrum} one should consider the limit $f_2\rightarrow \infty$ and $f_0\rightarrow 0$ in (\ref{rhopipi2}).
The trilinear coupling of $SU(2)_L$ heavy vectors to GBs simplifies in this limit as
\begin{equation}
\label{rhopipi}
\frac{g}{2}(\cot\theta_L-\tan\theta_L)\rho^\mu_{c_L}\partial_{\mu} \pi^{\widehat{a}}\pi^{\widehat{b}} f^{\widehat{a}\widehat{b}c_L}.
\end{equation}
Analogous fomulas hold for the other resonances. In our setup, contrary to \cite{contino-sundrum}, in the elementary composite basis
the Higgs couples to the elementary fields through nearest neighbor interactions producing
the additional (subleading) contribution proportional to $g\tan\theta_L$.

The effects of EWSB can be easily taken into account. Eq.~(\ref{cubic2}) still holds with the replacement
\begin{equation}
\begin{aligned}
f^{\widehat{a} \widehat{b}c'}&= -i~ {\rm Tr} \left[ [ T^{\widehat{a}}, T^{\widehat{b}}] \overline{T}^{c'} \right] \\
\overline{T}^{c'}&= e^{-i \frac {\overline \Pi}f} T^{c'} e^{i \frac {\overline \Pi}f},
\end{aligned}
\end{equation}
where the $\overline \Pi=\sqrt{2}\langle h \rangle T^{\widehat 4}$ is the Higgs VEV.
To see this we just have to notice that EWSB is equivalent to a rotation of the elementary fields.
As expected we find in that the EWSB is negligible for practical purposes.

\subsection{$\rho\to W^+_L W^-_L,\ Z_L h$ Decay Widths}

Composite resonances decay mostly into longitudinal SM gauge bosons and third generation quarks
which are the SM states coupled more strongly to the composite sector. By using the equivalence theorem
the decay into longitudinal SM gauge bosons are equal to leading order to the decay into the GBs.
This allows to use (\ref{rhopipi}) to compute the partial decay widths of the heavy neutral resonances in $W_L^+W_L^-$ and $Z_Lh$.
Two examples of neutral gauge bosons decay (with masses around 2 TeV), are given in the tables below.
\begin{table}[h]
\begin{center}
\begin{tabular}{|c|c|c|c|}
\hline
& $~~~\rho_{3L}~~~$ & $~~~\rho_{3R~~~}$ & $\rho_{X}$ \\ \hline
$m$(TeV) & 2.5  & 2.5 &  2.1 \\ \hline
$\Gamma_{W^+_LW^-_L}=\Gamma_{Z_Lh}$ (GeV) & 50 & 50 & $4\cdot 10^{-3}$ \\ \hline
\end{tabular}
\caption{\small  Decay widths of neutral resonances into $W^+_L W^-_L$ and $Z_L h$. $g_\rho=7$, $g_{\rho_X}=6$ and $f=0.5\ {\rm TeV}$.}
~\\
\begin{tabular}{|c|c|c|c|}
\hline
& $~~~\rho_{3L}~~~$ & $~~~\rho_{3R~~~}$ & $\rho_{X}$ \\ \hline
$m$(TeV) & 2.9  & 2.8 &  1.4 \\ \hline
$\Gamma_{W^+_LW^-_L}=\Gamma_{Z_Lh}$ (GeV) & 18 & 18 & $ 6\cdot 10^{-3}$ \\ \hline
\end{tabular}
\caption{\small  Decay widths of neutral resonances into $W^+_L W^-_L$ and $Z_L h$.  $g_\rho=4$, $g_{\rho_X}=2$ and $f=1\ {\rm TeV}$.}
\end{center}
\end{table}
In absence of mixing with the elementary fields the resonance associated to $U(1)_X$ does not decay into longitudinal SM gauge bosons.
A very small decay width is induced by the mixing, the coupling of $\rho_X$ to $W/Z$ is of the order of $g_{0Y}(g_{0Y}/g_\rho)$.
Before EWSB the decay of $\rho_{3L}$ and $\rho_{3R}$ is as expected identical in the two channels.
After EWSB the neutral states mix and the exact mass eigenstates decay unequally in $W^+_LW^-_L$ and $Z_Lh$.
However since the mass difference is very small it is safe to neglect this effect for practical purposes.
While $\rho_{3L}$ and $\rho_{3R}$ are almost degenerate in our general setup we do not expect $\rho_{X}$
to be closely degenerate. As a consequence it will not have a large mixing with the others even after EWSB and its decay remains
very small as found in the numerical examples above. This state can then only significantly decay into third generation quarks. For the same reason the production through vector boson fusion is suppressed and proceeds through Drell-Yan process.
The latter may be important even for $\rho_{3L}$ and $\rho_{3R}$ whose coupling to $W$, $Z$ is not suppressed.
This result is in general different from the one of 5D models where the resonances are approximately degenerate,
see \cite{Agashe:2007ki}.

\end{document}